\documentclass[a4paper, 12pt]{article}
\usepackage[a4paper]{geometry}
\usepackage{algorithm2e}
\usepackage{amsmath}
\usepackage{amssymb}
\usepackage{enumerate}
\usepackage{latexsym}
\usepackage{graphics}
\usepackage{graphicx,epstopdf}
\usepackage{subfigure}
\usepackage{caption}
\usepackage{wrapfig}
\usepackage{lineno}
\usepackage{stmaryrd}
\usepackage[english]{babel}
\usepackage{amsmath,amssymb,amsthm}
\usepackage{subfigure}
\graphicspath{{figures/}}
\usepackage{tikz}
\usetikzlibrary{shapes,arrows,backgrounds,snakes}
\tikzstyle{wide}=[draw, minimum size=2em, text width=7.5em, text
centered] 
\tikzstyle{narrow}=[draw, minimum size=2em, text
width=2em, text centered]

\newcommand{\be}{\begin{equation}}
\newcommand{\ee}{\end{equation}}
\newcommand{\bea}{\begin{eqnarray}}
\newcommand{\eea}{\end{eqnarray}}

\newcommand{\nod}{\noindent}

\newcommand{\ba}{\begin{array}}
\newcommand{\ea}{\end{array}}
\newcommand{\bc}{\begin{center}}
\newcommand{\ec}{\end{center}}

 \allowdisplaybreaks

\begin{document}
%\linenumbers*[1]
%\title{\bf Exact and approximate epidemic models on networks: a new, improved closure relation\\}
\title{\bf Higher-order structure and epidemic dynamics in clustered networks\\}
\author{Martin Ritchie$^{1}$, Luc Berthouze$^{2, 3}$, Thomas House$^{4}$ \& Istvan Z. Kiss$^{1,\ast}$}

\maketitle

\begin{center}

$^1$School of Mathematical and Physical Sciences, Department of
Mathematics, University of Sussex, Falmer,
Brighton BN1 9QH, UK\\
$^2$ Centre for Computational Neuroscience and Robotics, University of Sussex, Falmer, Brighton BN1 9QH, UK\\
$^3$ Institute of Child Health, London, University College London, London WC1E 6BT, UK\\
$^4$ Warwick Mathematics Institute, University of Warwick, Gibbet Hill Road, Coventry, CV4 7AL, UK

\end{center}

\vspace{12cm}
\begin{flushleft}
$\ast$corresponding author\\
email: i.z.kiss@sussex.ac.uk\\
\end{flushleft}

\newpage

\begin{abstract}
Clustering is typically measured by the ratio of triangles to all triples, open or closed. Generating clustered networks, and how clustering affects dynamics
on networks, is reasonably well understood for certain classes of networks \cite{vmclust, karrerclust2010}, e.g., networks composed of lines and non-overlapping 
triangles. In this paper we show that it is possible to generate networks which, despite having the same degree distribution and equal clustering, exhibit different higher-order structure, specifically, overlapping triangles and other order-four (a closed network 
motif composed of four nodes) structures. To distinguish and quantify these additional structural features, we develop a new network metric capable of measuring order-four structure which, when used alongside traditional network metrics, allows us to more 
accurately describe a network's topology. Three network generation algorithms are considered: a modified configuration model and two rewiring algorithms. By generating homogeneous networks with equal clustering we study and quantify their structural 
differences, and using SIS (Susceptible-Infected-Susceptible) and SIR (Susceptible-Infected-Recovered) dynamics we investigate computationally how differences in higher-order structure impact on epidemic threshold, final epidemic or prevalence levels and 
time evolution of epidemics. Our results suggest that characterising and measuring higher-order network structure is needed to advance our understanding of the impact of network topology on dynamics unfolding on the networks.
\end{abstract}

\nod {\bf Keywords:} Clustering, motif counting, loops, networks, epidemic.

\newpage

%%%%%%%%%%%%%%%%%%%%%%%%%%%%%%%%%%%%%%%%%%%%%%%%%%%%%%%%%%%%%%%%%%%%%%%%%%%%%%%%%%%%%%%%%%
\section{Introduction}
%%%%%%%%%%%%%%%%%%%%%%%%%%%%%%%%%%%%%%%%%%%%%%%%%%%%%%%%%%%%%%%%%%%%%%%%%%%%%%%%%%%%%%%%%%

Network modelling is an essential tool in characterising a wide range of phenomena: infectious diseases, brain activity, chemical reactions, social interactions, the internet, etc. Any system that involves
interactions of its constituent components may be modelled as a network. The versatility of networks as a modelling paradigm may be further augmented by running dynamical processes on the network 
such as epidemics or neuronal activity. A network's structure can have a dramatic effect on the processes that run on the network which is currently parameterised by low-order structure alongside
the degree distribution. As we shall see with epidemiological processes the presence of higher-order structure affects how a disease spreads through a network, and the effect of such structures on neuronal dynamics 
is known to be significant \cite{sporns2005human, honey2009predicting, gallos2012small, lynall2010functional, kaiser2010optimal}. 
%In this paper we define a motif or network structure to be a group of nodes that are connected in a specific way e.g. loops, complete structures, stars or lollipops. 
%An order-four structure is any structure that is contained within a closed loop of four nodes and we refer to structures with five or more nodes as higher-order structure.  
In this paper we aim to go beyond open and closed triples and give a more comprehensive description of networks in terms of higher-order structure frequency (specifically order-four structures) and their distribution around nodes. 
In particular, we will examine existing clustered network generating algorithms with respect to their ability, or otherwise, to control higher-order network structure which sometimes may be regarded as 
a by-product of generating low-order structure that can preclude a correct interpretation of the impact of clustering. 
The paper is structured as follows. We first introduce and describe a set of clustered network generating algorithms. We follow with a presentation of the network metrics (including a description of the motif identifying/counting algorithm) that we propose to quantify similarities and differences between the generated networks. We then analyse and discuss the impact of higher-order structural differences, at identical degree distribution and equal clustering, on SIS and SIR epidemics. Finally, we discuss how our motif-counting results and newly proposed measure for higher-order structures could be used to parameterise pairwise-like models with closure at the level of quadruples.  

%%%%%%%%%%%%%%%%%%%%%%%%%%%%%%%%%%%%%%%%%%%%%%%%%%%%%%%%%%%%%%%%%%%%%%%%%%%%%%%%%%%%%%%%%%
\section{Material and methods}
%%%%%%%%%%%%%%%%%%%%%%%%%%%%%%%%%%%%%%%%%%%%%%%%%%%%%%%%%%%%%%%%%%%%%%%%%%%%%%%%%%%%%%%%%%

%%%%%%%%%%%%%%%%%
\subsection{Network construction}
%%%%%%%%%%%%%%%%%
A significant part of network research relies on networks with arbitrary degree distributions
built using the configuration model \cite{newman2009random}. 
This algorithm generates networks where nodes mix at random and where the probability that two nodes are connected is simply proportional to the product of their degree.  Such networks coupled with 
stochastic node dynamics such as SIS, SIR or neural dynamics, are amenable 
to developing macroscopic low-dimensional ODE models that are in excellent agreement with values obtained 
from stochastic simulations. By construction, these networks are loop-less in the limit of large network size.
While such networks can be considered in many cases as realistic or plausible models of some real-world networks, there are many
instances where networks have a high degree of structure that typically involves clusters of well connected nodes. 
Classic examples come from household models used in epidemiology \cite{ball2010analysis, ball2001stochastic}, and networks of social interactions in general.
Motivated by this, there are a series of theoretical or synthetic network models that can be tuned to display increased levels of clustering \cite{vmclust, karrerclust2010, newman2009random, read2003disease, eames2008modelling, bansal2009exploring}, 
where clustering denotes the ratio of closed loops of length three with respect to all possible open triple, irrespective of whether they are closed or not.

The classic algorithms to generate networks with tunable clustering include: (a) the spatial algorithm by Reed et al. \cite{read2003disease}, 
(b) an iterative method proposed by Eames \cite{eames2008modelling}, (c) a configuration model that includes clustering \cite{karrerclust2010} and
(d) the Big-V rewiring algorithm \cite{bansal2009exploring, housetracing}. In a recent study, Green et al. \cite{green2010large} showed that even under identical degree 
distributions and equal levels of clustering, networks built based on different algorithms can display a markedly different `higher-order structure'. While their analysis identified large scale structural differences amongst networks with identical degree distribution and clustering, it did not consider extending the concept of clustering involving three nodes to higher-order structures with four or more nodes.
The concept of motifs is not new \cite{sporns2005human,karrerclust2010, vmclust, Keeling22041999, house2011insights} and understanding network structure through higher-order motifs is going to provide 
a level of detail which cannot be articulated by open or connected triples alone. Below we provide a brief description of the clustered network construction algorithms used in this paper.

%%%%%%%%%%%%%%%%%%
\subsubsection{Big-V rewiring}
\begin{figure}
\centering
\begin{tikzpicture}[thick,inner sep=1mm]
\node (u) at ( 0.00, 0)   [circle,draw=blue!50,fill=blue!40] {$u$};
\node (v) at ( 0.75,-1)   [circle,draw=blue!50,fill=blue!40] {$v$};
\node (w) at ( 1.50,-2)   [circle,draw=blue!50,fill=blue!40] {$w$};
\node (x) at ( 2.25,-1)   [circle,draw=blue!50,fill=blue!40] {$x$};
\node (y) at ( 3.00, 0)   [circle,draw=blue!50,fill=blue!40] {$y$};
\draw[line width=1.25pt,color=red]  (u) -- (v);
\draw[line width=1.25pt,color=blue]  (v) -- (w);
\draw[line width=1.25pt,color=blue]  (w) -- (x);
\draw[line width=1.25pt,color=red]  (x) -- (y);
\draw [->] (3.50,-1.00) -- (5.50,-1.00);
\node (uu) at ( 6.00, 0)   [circle,draw=blue!50,fill=blue!40] {$u$};
\node (vv) at ( 6.75,-1)   [circle,draw=blue!50,fill=blue!40] {$v$};
\node (ww) at ( 7.50,-2)   [circle,draw=blue!50,fill=blue!40] {$w$};
\node (xx) at ( 8.25,-1)   [circle,draw=blue!50,fill=blue!40] {$x$};
\node (yy) at ( 9.00, 0)   [circle,draw=blue!50,fill=blue!40] {$y$};
\draw[line width=1.25pt,color=red]   (uu) -- (yy);
\draw[line width=1.25pt,color=red]   (vv) -- (xx);
\draw[line width=1.25pt,color=blue]  (ww) -- (xx);
\draw[line width=1.25pt,color=blue]  (vv) -- (ww);
\end{tikzpicture}
\caption{A single Big-V rewiring. (a) Identify a chain of 5 nodes with 4 edges and (b) if edges $(u$-$v)$ or $(x$-$y)$ are already part of a triangle the cuts will not be made, otherwise rewiring is performed, and (c) independently of the outcome of (b) the algorithm will proceed to find a new chain.}
\label{fig:bigv}
\end{figure}
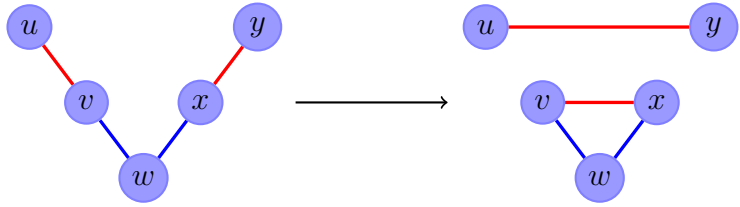

%%%%%%%%%%%%%%%%%%
The `Big-V' is an iterative rewiring algorithm that can introduce clustering into any given network and is commonly used by network scientists \cite{bansal2009exploring,
housetracing,green2010large}. At each iterative step, a chain of 5 distinct nodes ($u$-$v$-$w$-$x$-$y$) is selected at random and a clone network is generated where the links ($u$-$v$) 
and ($x$-$y$) are broken and the edges ($u$-$y$) and  $(v$-$x)$ are created. This leads to a single chain of 5 nodes being broken into a triangle and a disconnected pair, see Fig.~\ref{fig:bigv}. Local clustering for each node in the chain, as well as all of its neighbours, is computed in both the original and cloned networks and the new configuration is kept only if the level of clustering has increased.

%%%%%%%%%%%%%%%%%%%%%%%%%%%
\subsubsection{Motif decomposition rewiring}
\begin{figure}
 \centering
 \includegraphics[scale=0.35]{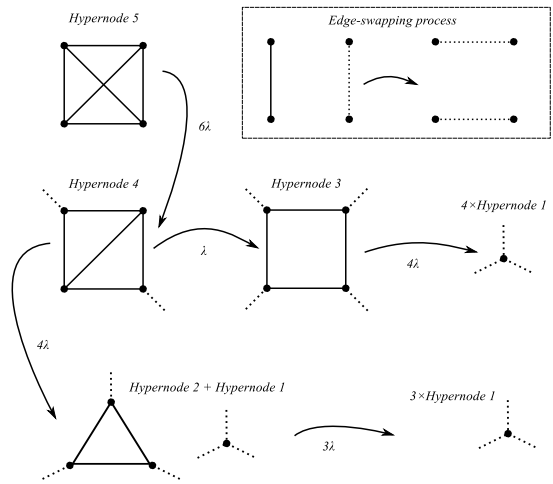} \captionof{figure}{MD hyper-node configurations. The different hypernode configurations of a homogeneous graph with k = 3 as edges are decomposed from local to global.}
 \label{fig:housedraft}
\end{figure}

%%%%%%%%%%%%%%%%%%%%%%%%%%%
MD (Motif Decomposition) is an iterative rewiring algorithm that starts with a collection of complete sub-networks that are disconnected from one another and rewires edges randomly to reduce the clustering from its maximal value of 1 to the desired level. The following steps are performed: 
\begin{enumerate}[i.]
  \item initialise a network that is composed of $m$ complete motifs each with $n$ members so that: $N=nm$ and $\langle k \rangle = n-1$,
  \item categorize every edge as `local',
  \item for the first step only, select at random two local edges, cut them, and swap the stubs to form new edges. Mark the pair of new edges as global,
  \item select a local and a global edge, cut them, and swap the stubs to form new edges. Mark the pair of new edges as global,
  \item check the global clustering, if the desired level has not been achieved repeat step (iv). 
\end{enumerate}
Fig.~\ref{fig:housedraft} illustrates this process being performed on a complete motif with 4 members. It should be noted that this method may work with a heterogeneous degree distribution in which case 
the network would need to be initialised with motifs of $k+1$ nodes for each different degree $k$. MD has the significant advantage that it is computationally cheap and that, in the limit of large networks, network properties can be calculated analytically (see appendix \ref{sec:mdanalysis}).
%%%%%%%%%%%%%%%%%%%%%%%%%%%
\subsubsection{CCM (Clustered Configuration Model)}
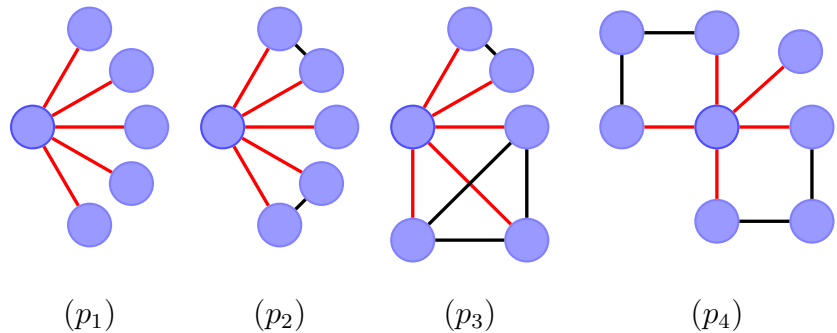
\begin{figure}
\centering
\begin{tikzpicture}[thick,inner sep=2mm]
\node (A)  at ( -7.5 ,   -4)    [circle,draw=blue!70,fill=blue!40] {};
\node (Aa) at ( -6.75,   -5.30) [circle,draw=blue!50,fill=blue!40] {};
\node (Ab) at ( -6.20,   -4.75) [circle,draw=blue!50,fill=blue!40] {};
\node (Ac) at ( -6.00,   -4.00) [circle,draw=blue!50,fill=blue!40] {};
\node (Ad) at ( -6.20,   -3.25) [circle,draw=blue!50,fill=blue!40] {};
\node (Ae) at ( -6.75,   -2.70) [circle,draw=blue!50,fill=blue!40] {};
\draw[line width=1.25pt,color=red]  (A) -- (Aa);
\draw[line width=1.25pt,color=red]  (A) -- (Ab);
\draw[line width=1.25pt,color=red]  (A) -- (Ac);
\draw[line width=1.25pt,color=red]    (A) -- (Ad);
\draw[line width=1.24pt,color=red]    (A) -- (Ae);
\node (B)  at ( -5.0 ,   -4)    [circle,draw=blue!70,fill=blue!40] {};
\node (Bd) at ( -4.25,   -2.70) [circle,draw=blue!50,fill=blue!40] {};
\node (Be) at ( -3.70,   -3.25) [circle,draw=blue!50,fill=blue!40] {};
\node (Bc) at ( -3.50,   -4.00) [circle,draw=blue!50,fill=blue!40] {};
\node (Ba) at ( -3.70,   -4.75) [circle,draw=blue!50,fill=blue!40] {};
\node (Bb) at ( -4.25,   -5.30) [circle,draw=blue!50,fill=blue!40] {};
\draw[line width=1.25pt,color=red]  (B) -- (Ba);
\draw[line width=1.25pt,color=red]  (B) -- (Bb);
\draw[line width=1.25pt,color=red]  (B) -- (Bc);
\draw[line width=1.25pt,color=red]    (B) -- (Bd);
\draw[line width=1.24pt,color=red]    (B) -- (Be);
\draw[line width=1.25pt,color=black]  (Bb) -- (Ba);
\draw[line width=1.25pt,color=black]  (Bd) -- (Be);
\node (C)  at (  -2.5, -4)   [circle,draw=blue!70,fill=blue!40] {};
\node (Ca) at ( -1.20, -3.25)[circle,draw=blue!50,fill=blue!40] {};
\node (Cb) at ( -1.75, -2.70)[circle,draw=blue!50,fill=blue!40] {};
\node (Cc) at ( -1   , -4)   [circle,draw=blue!50,fill=blue!40] {};
\node (Cd) at ( -1   , -5.5) [circle,draw=blue!50,fill=blue!40] {};
\node (Ce) at ( -2.5 , -5.5) [circle,draw=blue!50,fill=blue!40] {};
\draw[line width=1.25pt,color=red]  (C) -- (Ca);
\draw[line width=1.25pt,color=red]  (C) -- (Cb);
\draw[line width=1.25pt,color=black] (Ca) -- (Cb);
\draw[line width=1.25pt,color=red] (C) -- (Cc);
\draw[line width=1.25pt,color=red] (C) -- (Cd);
\draw[line width=1.24pt,color=red] (C) -- (Ce);
\draw[line width=1.25pt,color=black] (Cc) -- (Cd);
\draw[line width=1.25pt,color=black] (Cc) -- (Ce);
\draw[line width=1.24pt,color=black] (Cd) -- (Ce);
\node (D) at  (  1.5 ,   -4)   [circle,draw=blue!70,fill=blue!40] {};
\node (Da) at (  1.5 ,   -2.75)[circle,draw=blue!50,fill=blue!40] {};
\node (Db) at (  0.25,   -4)   [circle,draw=blue!50,fill=blue!40] {};
\node (Dc) at (  0.25,   -2.75)[circle,draw=blue!50,fill=blue!40] {};
\node (Dd) at (  2.75,   -4)   [circle,draw=blue!50,fill=blue!40] {};
\node (De) at (  2.75,   -5.25)[circle,draw=blue!50,fill=blue!40] {};
\node (Df) at (  1.5 ,   -5.25)[circle,draw=blue!50,fill=blue!40] {};
\node (Dg) at (  2.6 ,   -3.0) [circle,draw=blue!50,fill=blue!40]{};
\draw[line width=1.25pt,color=red]  (D) -- (Da);
\draw[line width=1.25pt,color=red]  (D) -- (Db);
\draw[line width=1.25pt,color=black] (Da) -- (Dc);
\draw[line width=1.25pt,color=black] (Db) -- (Dc);
\draw[line width=1.25pt,color=red] (D) -- (Dd);
\draw[line width=1.24pt,color=red] (D) -- (Df);
\draw[line width=1.25pt,color=black] (De) -- (Df);
\draw[line width=1.25pt,color=black] (Dd) -- (De);
\draw[line width=1.25pt,color=red] (D) -- (Dg);
\node (labelA) at  ( -6.75, -6.5) [] {($p_1$)};
\node (labelB) at  ( -4.25, -6.5) [] {($p_2$)};
\node (labelC) at  ( -1.75, -6.5)[] {($p_3$)};
\node (labelD) at  (  1.5, -6.5) [] {($p_4$)};
\end{tikzpicture}
\caption{Corner/edge allocation. A node is initially allocated a quintuple of stubs. With probability $p_1,~p_2,~p_3$ and $p_4$ the node will be part of a number of different structures as shown above. In this work homogeneous
networks with $\langle k \rangle = 5$ have been used. If a different degree or degree distribution is required then the configuration of motifs will need to be adjusted accordingly.}
 \label{fig:ccmfig}
\end{figure}

%%%%%%%%%%%%%%%%%%%%%%%%%%%
It is possible to modify the configuration model so that it constructs networks using specified motifs. Karrer et al. \cite{karrerclust2010} 
and Volz et al. \cite{vmclust} have shown how to build networks using a configuration model that includes triangle motifs.
This idea may be easily extended to allow for larger and more exotic motifs to be included in the networks' construction.  
Rather than just lines, the number of lines and corners of motifs that originate from a node can be varied. 
In any given motif a node can be considered as a corner and the number of stubs originating from this node that join it to the motif defines its corner type, 
essential in describing corners of asymmetric structures. To generate a network using this method, the following steps are performed:
\begin{enumerate}
 \item allocate to a node a number of stubs following a given degree distribution,
 \item multinomially determine the configuration of corners and single stubs, 
 \item create lists for each corner type where a node that is allocated $\kappa$ corners of a certain type, will appear $\kappa$ times in the corresponding corner list,  
 \item draw corners at random and without replacement from the appropriate lists and connect with other corners to form motifs, 
 \item repeat until all lists are empty.   
\end{enumerate}
Fig.~\ref{fig:ccmfig} illustrates corner allocation for an example node. Due to the nature of the configuration model self loops and double loops may be formed. The expected number of such occurrences depends only on degree, becoming negligibly small in the limit of large networks \cite{newman2009networks}.  

In this paper homogeneous CCM networks are used with clustering of $\phi = 0.2$ and $\phi= 0.4$. The stub configurations to generate such networks are as follows:
\begin{enumerate}
 \item $\phi = 0.2$: with probability $p_1=0.5$ the quintuple of stubs is maintained as independent links, and with probability $p_2=1-p_1$ the quintuple is arranged into one complete square corner and one triangle corner,
 \item $\phi = 0.4$: every node is allocated one complete square corner and one triangle corner,
 \item $\phi = 0.8$: as this algorithm does not allow overlaps between motifs, this value of clustering cannot be achieved.  
\end{enumerate}
\begin{table}
\begin{center}
    \begin{tabular}{l|l|l|l}
    \hline
    ~            & Lines & Triangles & Complete squares \\ \hline
    $\phi = 0.2$ & 2.5   & 0.5       & 0.5              \\ \hline
    $\phi = 0.4$ & 0     & 1         & 1                \\
    \end{tabular}
\end{center}
\caption{The expected number of lines, triangles and complete squares per node for each level of clustering used.}
\label{table:tab2}
\end{table} 
Table \ref{table:tab2} shows the expected motif allocation per node. 
The configuration model allow us to analytically determine some of the networks structural properties, more specifically the PGF (Probability Generating Function) of the degree/motif distribution.

The CCM algorithm for this work was configured as follows. First, initialise each node with five stubs. Then let $p_1$ denote the probability that the five stubs form lines, $p_2$ two triangles and one line, 
$p_3$ one complete square and one triangle and $p_4$ two empty squares and one line. The probabilities are chosen such that $\sum_ip_i=1$. Let $x_i$ denote the dummy variables of the PGF that corresponds
to corner types: $x_1$ (simple stubs), $x_2$ (two triangles and a simple stub), $x_3$ (a complete square and a triangle), $x_4$ (two empty squares and a simple stub). The PGF of the networks degree/corner distribution may now be written as:
\begin{eqnarray}
 \Psi(x_1,x_2,x_3,x_4) = x_1^5p_1 + x_1x_2^2p_2 + x_2x_3p_3 + x_1x_4^2p_4,
\end{eqnarray}
and the original stub distribution may be recovered by substituting each $x_i$ with $x_1^n$ where $n$ is the corner-stub cardinality:
\begin{eqnarray}
 \psi(x_1)&=& x_1^5p_1 + x_1(x_1^2)^2p_2 + x_1^2 x_1^3 p_3 + x_1(x_1^2)^2p_4 \\
	&=& x_1^5(p_1 + p_2 + p_3 + p_4) = x_1^5.
\end{eqnarray}
$\psi'(1)$ yields the expected degree, and $N\psi''(1)/2$ yields the number of paths of length three in the network
\cite{vmclust}. The number of unique triangles in the network can be determined by $\Psi$: 
\begin{eqnarray}
[\triangle] = N\left(\frac{\Psi_{x_2}(1,1,1,1)}{3}+ \frac{4\cdot\Psi_{x_3}(1,1,1,1)}{4}\right),
\end{eqnarray}
since each square is quadruply counted and contains four separate triangles. Clustering is measured as the ratio of three times the number of triangles to all closed and unclosed triples:
\begin{eqnarray}
\phi_{global} &=& \frac{3N\left(\frac{\Psi_{x_2}(1,1,1,1)}{3}+ \Psi_{x_3}(1,1,1,1)\right)}{N\psi''(1)} \\ 
	      &=& \frac{\Psi_{x_2}(1,1,1,1)+ 3\Psi_{x_3}(1,1,1,1)}{\psi''(1)} \\
				&=& \frac{p_2 + 2p_3}{5} 
\end{eqnarray} 
For the two types of CCM networks used in this study: $p_1=0.5,~p_3=0.5$ yields $\phi = 0.2$ and $p_3=1$ yields $\phi=0.4$ (see table \ref{table:tab2}). 
%%%%%%%%%%%%%%%%%%%%%%%%%%%%%%%%%%%%%%%%%%
\subsection{Network metrics: Third and higher-order network structure}
%%%%%%%%%%%%%%%%%%%%%%%%%%%%%%%%%%%%%%%%%%
Here we give a succinct summary of the classic and newly proposed network metrics that will be used to
compare and contrast the networks resulting from the different algorithms. Although the novelty of the paper is around order-four structure, we will first consider classic (or third-order) network measures, 
such as clustering in the global sense as well as distribution of clustering at node level, nodal betweenness centrality, and
connected component analysis via percolation.  We then augment the classic network descriptions with an analysis of the distribution of motifs of order higher than closed and open triples both globally and on a per node basis. A network of $N$ individuals is represented with an adjacency matrix, $A \in \{0,1\}^{N^2}$. A pair of individuals $(i,j)$ share a connection if $A_{i,j}=1$. The networks are undirected, $A = A^T$, and self loops are
not allowed $A_{i,i}=0,~ \forall i \in N$.  

\begin{enumerate}
\item \textbf{Clustering:} clustering may be defined in two ways \cite{watts1998collective}: Local (node level) and global (network level). The local clustering of a node $n$, of degree $n_k$, is the ratio of connections between neighbours of $n$ and potential connections of neighbours of $n$. Let $\mathcal{N}$ denote the sub-adjacency matrix of the neighbourhood of $n$ then:
\begin{eqnarray}
 \phi_{local} = \frac{\sum_{i,j}\mathcal{N}_{i,j}/2}{n_k(n_k-1)/2}.
\end{eqnarray}
Global clustering is defined as the ratio of the total number of closed triples to the total number of connected structures with 3 nodes. This may be computed from the adjacency matrix as \cite{keeling1999effects}:
\begin{eqnarray}
 \phi_{global} = \frac{trace(A^3)}{\|A^2\|-trace(A^2)},
\end{eqnarray}
where $\| A^2\|$ denotes the sum of all elements of $A^2$. Manipulating the adjacency matrix in this way yields multiplicative counts. An alternative method to obtain the equivalent counts is as follows: 
\begin{eqnarray}
  [\vee + \triangle] = \sum_{i,j,k,~i\neq j \neq k}a_{i,j}a_{j,k},
 \end{eqnarray}
yielding all connected structures of 3 nodes (closed and unclosed), similarly
 \begin{eqnarray}
[\triangle] = \sum_{i,j,k,~i\neq j \neq k}a_{i,j}a_{i,k}a_{j,k},
 \end{eqnarray}
yielding six times the number of unique triangles. A more complete description of this approach is provided in Appendix~\ref{sec:unique}, along with a conjecture of a possible mapping between unique and multiplicative counts. 

\item \textbf{Nodal betweenness centrality:} Nodal betweenness centrality measures how often a node appears in the set of shortest paths (which we shall denote $s$), geodesics, of the network \cite{freeman1977set}. 
Nodes with high betweenness centrality will more frequently appear in shortest paths than low ranked nodes. The betweenness centrality of a node $n$ can be computed by:
\begin{eqnarray}
 B_{bc}(n) = \frac{\sum_{i\neq j\neq n}s_{i,j}(n)}{|s_{i,j}|},
\end{eqnarray}
 where $s_{i,j}(n)$ denotes the number of shortest paths from $i$ to $j$ that contain node $n$. The removal of nodes with high betweenness centrality can significantly affect the flow of dynamical processes on the network \cite{newman2009networks}. 

\item \textbf{Connected component analysis:} CCs (connected components) are sets of nodes where any node may be reached from any other node that is a member of the set. CCs are used to describe the macroscopic structure of a network,
as opposed to clustering which describes the local structure of the network. Highly clustered networks contain many components that are weakly connected to, or disconnected from, one another. 
It has previously been shown \cite{green2010large} that the GCCs (Giant Connected Component) of highly clustered networks are composed of many CCs of varying size and that removing a low proportion of edges can be enough to isolate parts of the network. To perform the analysis: we generate a list of all edges in a network, cycle through each edge in the list and remove it with probability $p_r$, compute the size and frequency of all components remaining, and plot the cumulative distribution of component size.  

\item \textbf{Motif frequency and distribution:} Clustering (local or global) essentially measures the occurrence of triangles in a network. It does not distinguish two separate triangles from two triangles that share an edge, 
neither can it describe loops of order-four or larger. From the perspective of characterising higher-order structure it is a very coarse measurement. In this paper all closed structures of order-four  i.e. empty 
squares, diagonal squares, and complete squares motifs are considered at both network and node levels. It is possible to define new clustering type metrics using structures larger than triangles. Proceeding in a way similar to classic (third-order) clustering and limiting ourselves to 4-node structures connected in a loop, it is possible to define four new structural measurements:
the ratio of unclosed quadruples ($1 - \phi^1_4$), `empty' squares ($\phi^2_4$), squares with a single diagonal ($\phi^3_4$), and complete squares ($\phi^4_4$) to all connected structures of 4 nodes. We present our results in two formats: 
($i$) global ratios of {\em unique} order-four structure counts to all unique paths counts, closed and unclosed. 
($ii$) probability distribution of finding $x$ structures of a certain type associated with a given node. 
These measurements alongside clustering will provide a higher resolution analysis of network architecture. 
A brief synopsis of how to compute non-trivial paths of length $l+1$  is as follows (see appendix \ref{sec:algorithm} for the full pseudo-code, all path lengths refer to the number of {\em edges} and $A(\cdot,\cdot)$ 
is the adjacency matrix):
\begin{enumerate}
 \item consider a path $P$ of length $l$, and identify a head $H(P)$ ($1^{st}$ node of the path) and a tail $T(P)$ (the last node).
 \item For each neighbour $n$ of $T(P)$, if (i) $A(H(P),n)=1$, (ii) it has not already been counted as a closed path, and (iii) its reverse has not been counted as a closed path then count a closed path of length $l+1$, 
 \item for each neighbour $n$ of $T(P)$, if (i) $A(H(P),n)=0$, (ii) it has not already been counted as an open path, and (iii) its reverse has not been counted as an open path then count an open path of length $l+1$,
 \item for all closed paths of length $l+1$ remove circular and reverse circular permutations, 
 \item categorize each closed path by its completeness, i.e., the number (if any) of diagonals in a square.
\end{enumerate}
\end{enumerate}

%%%%%%%%%%%%%%%%%%%%%%%%%%%%
\subsection{Dynamics on networks}
%%%%%%%%%%%%%%%%%%%%%%%%%%%%
To establish the overall impact of higher-order network structure, simulations of various dynamics are performed on the generated networks.
First, we use the Markovian SIR (susceptible-infected-recovered) model with a per-contact infection rate $\tau$ and recovery rate $\gamma$. 
All simulations are performed using the Gillespie algorithm \cite{gillespie1976general}. To assess the impact of loops and cycles, we also simulate the SIS 
epidemic which is more likely to highlight differences in the cycle/motif composition. We shall see that structural differences between networks with the same degree distribution and clustering
manifest in epidemiological differences with regard to dynamics on the networks. Previous work \cite{wrap5586} used Kirkwood's superposition approximation to predict the effect of order-four structure
on epidemic dynamics. For SIS dynamics it was concluded that the presence of empty square structures reduces the endemic state for all levels of $\phi$, complete squares may increase or decrease the endemic state, 
and that diagonal squares had very little effect on the endemic state. In the following we make comparisons between networks that use different distributions of order-four motifs. We expect that networks with markedly different order-four motif distributions produce different epidemiological behaviour.

%%%%%%%%%%%%%%%%%%%%%%%%%%%%%%%%%%%%%%%%%%%%%%%%%%%%%%%%%%%%%%%%%%%%%%%%%%%%%%%%%%%%%%%%%
\section{Results}
%%%%%%%%%%%%%%%%%%%%%%%%%%%%%%%%%%%%%%%%%%%%%%%%%%%%%%%%%%%%%%%%%%%%%%%%%%%%%%%%%%%%%%%%%
Using the various construction algorithms, we give an overarching analysis of structural differences between networks with the same degree distribution and same levels of classic clustering.
All networks used are homogeneous with $\langle k \rangle=5$, allowing for the formation of structures/loops while keeping the
complexity to a manageable level. We carry out our analysis on a range of clustering values (i.e. $\phi=0.2, 0.4, 0.8$) to measure and evaluate the extent to which clustering can emerge from, or determine, different configurations of order-four structures.

\begin{figure}
\centering
\begin{tabular}{ccc}
 \includegraphics[scale=0.45]{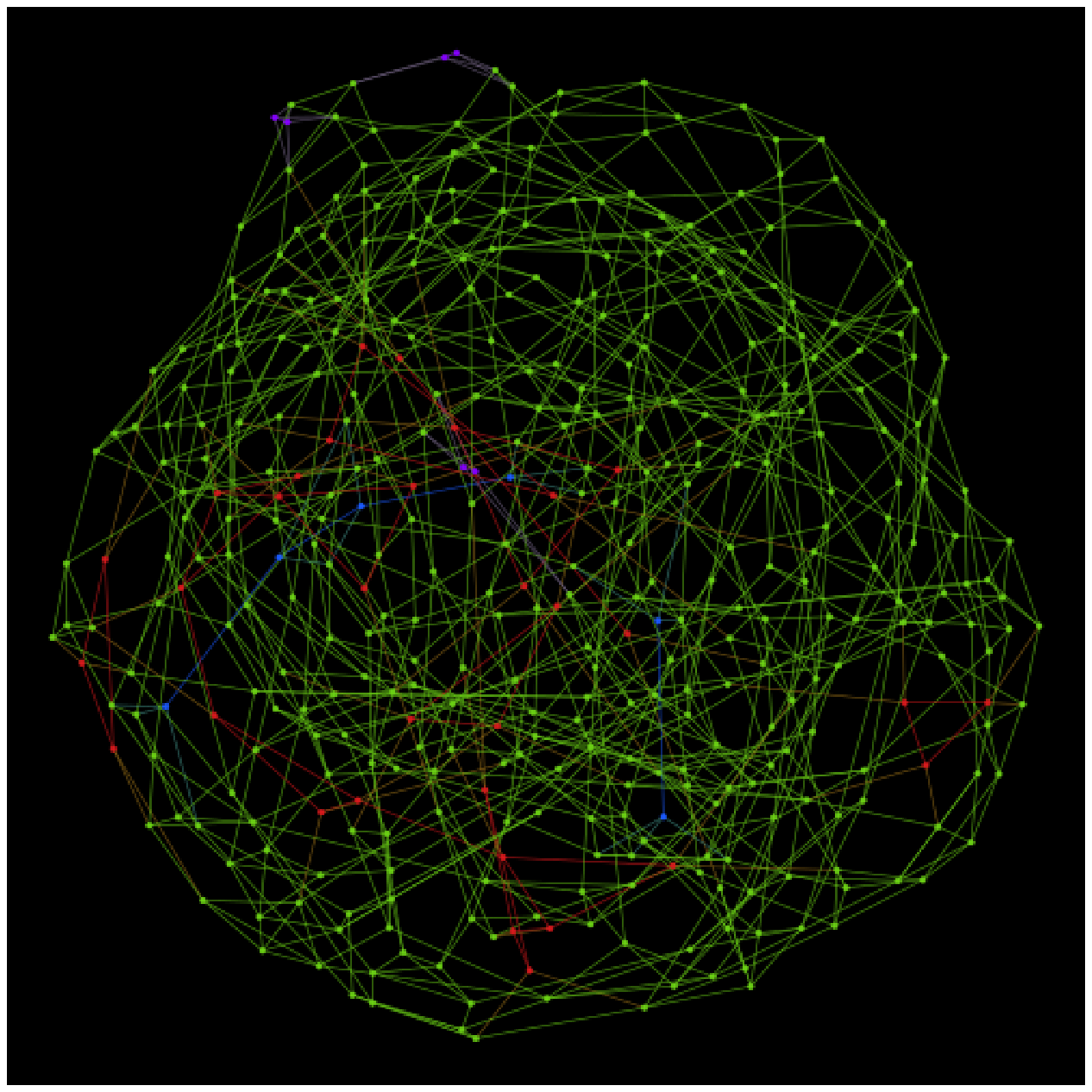} & 
 \includegraphics[scale=0.45]{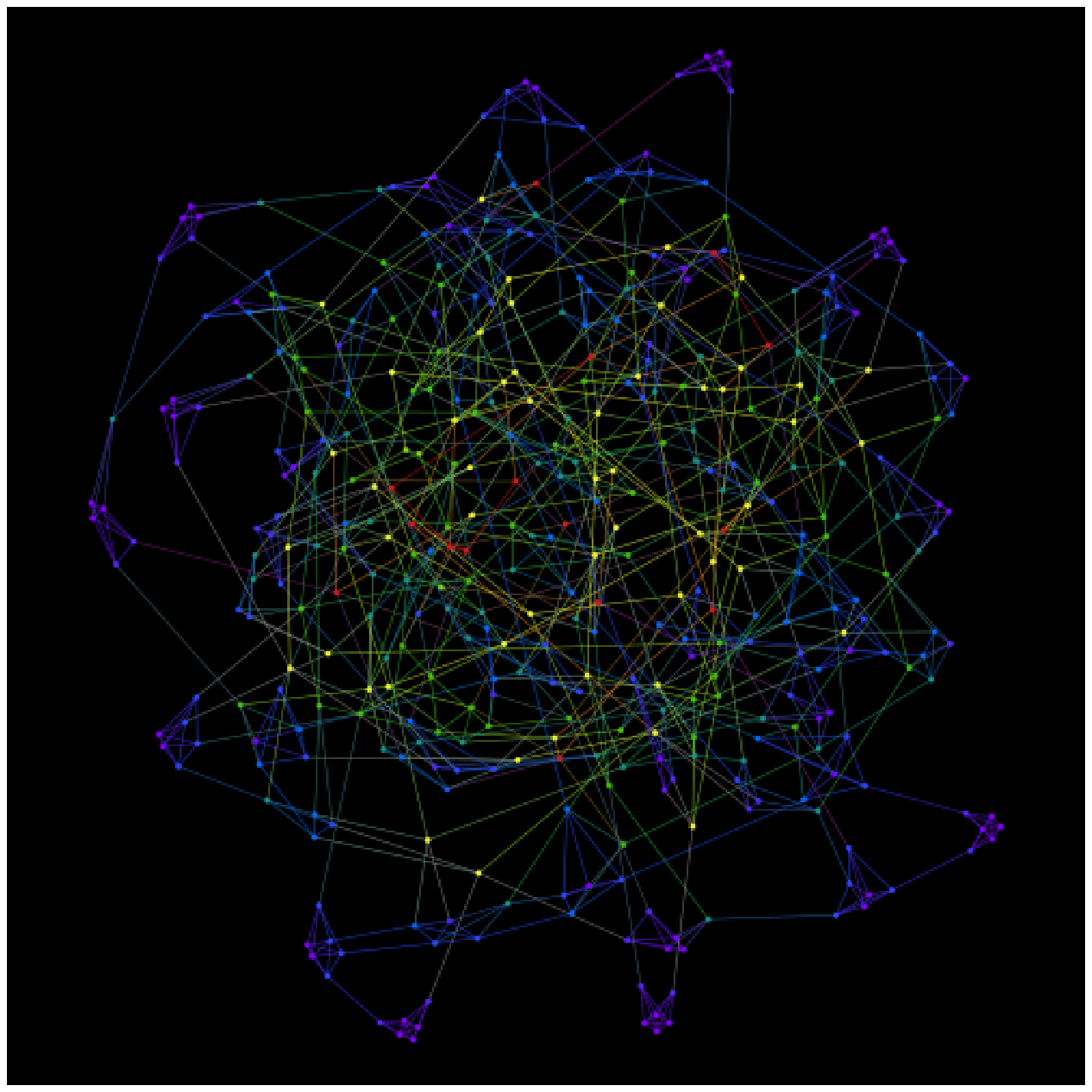} &
 \includegraphics[scale=0.45]{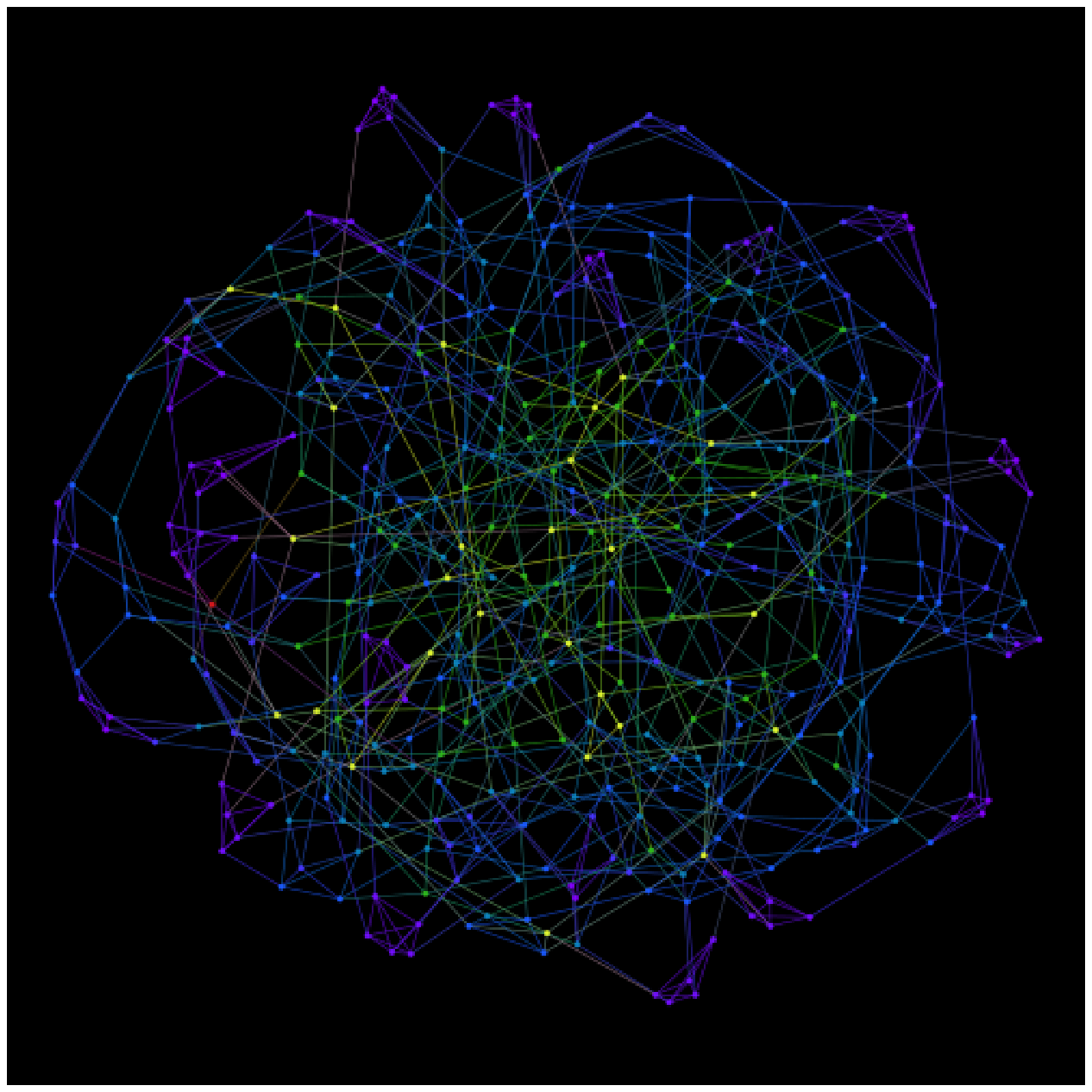} \\
CCM & MD & Big-V \\
\end{tabular}
\caption{Example networks. Homogeneous networks with all parameters held equal, $N= 400$, $\langle k \rangle = 5$ and $\phi = 0.4$. The nodes are coloured so those at the red end of the spectrum 
have low local clustering, and those at the blue end of the spectrum have high local clustering.}
\label{fig:gephi1}
\end{figure}

\begin{figure}
\minipage{0.32\textwidth}
  \includegraphics[width=\linewidth]{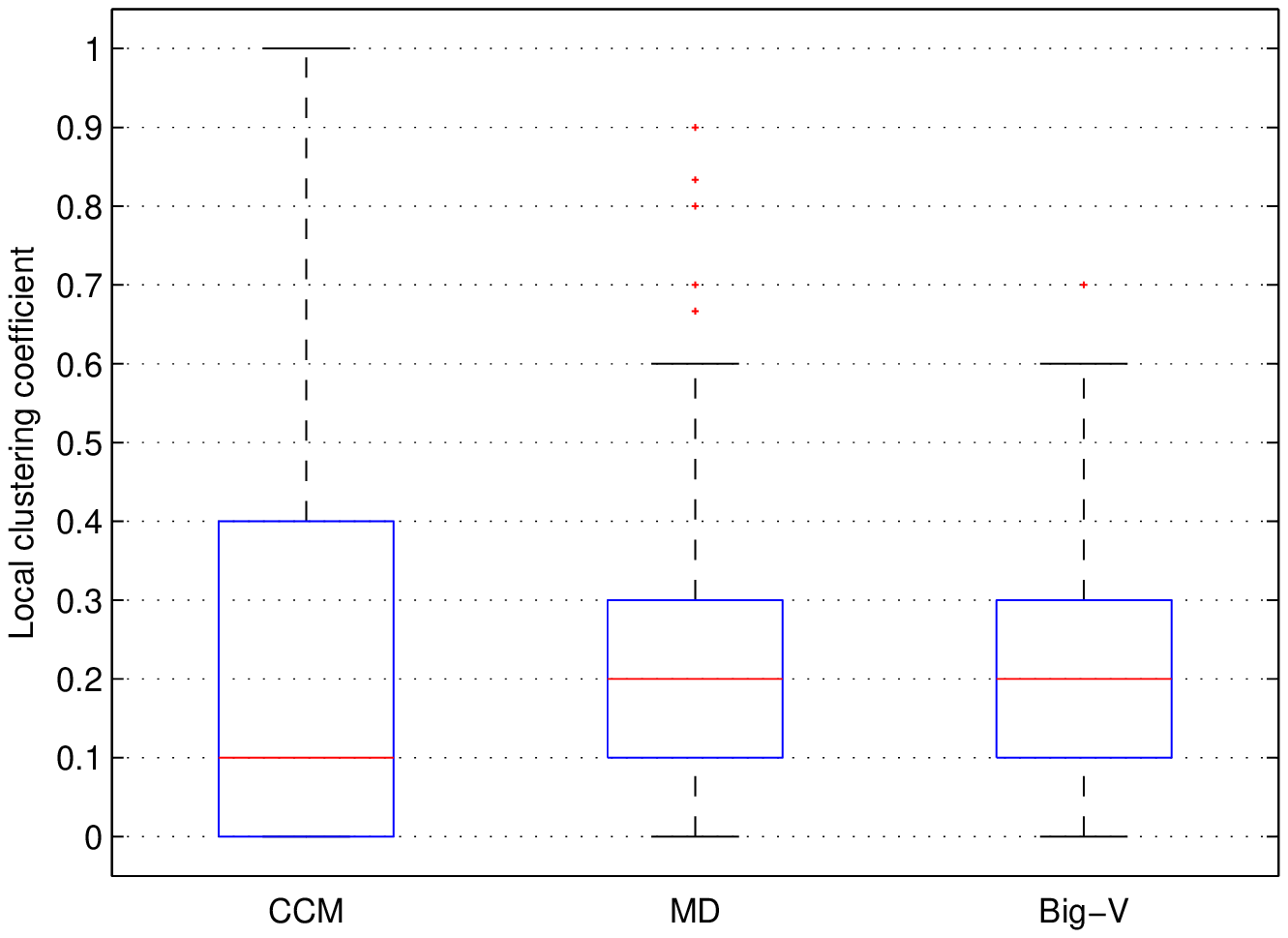}
  \caption*{$\phi =  0.2$}
\endminipage\hfill
\minipage{0.32\textwidth}
  \includegraphics[width=\linewidth]{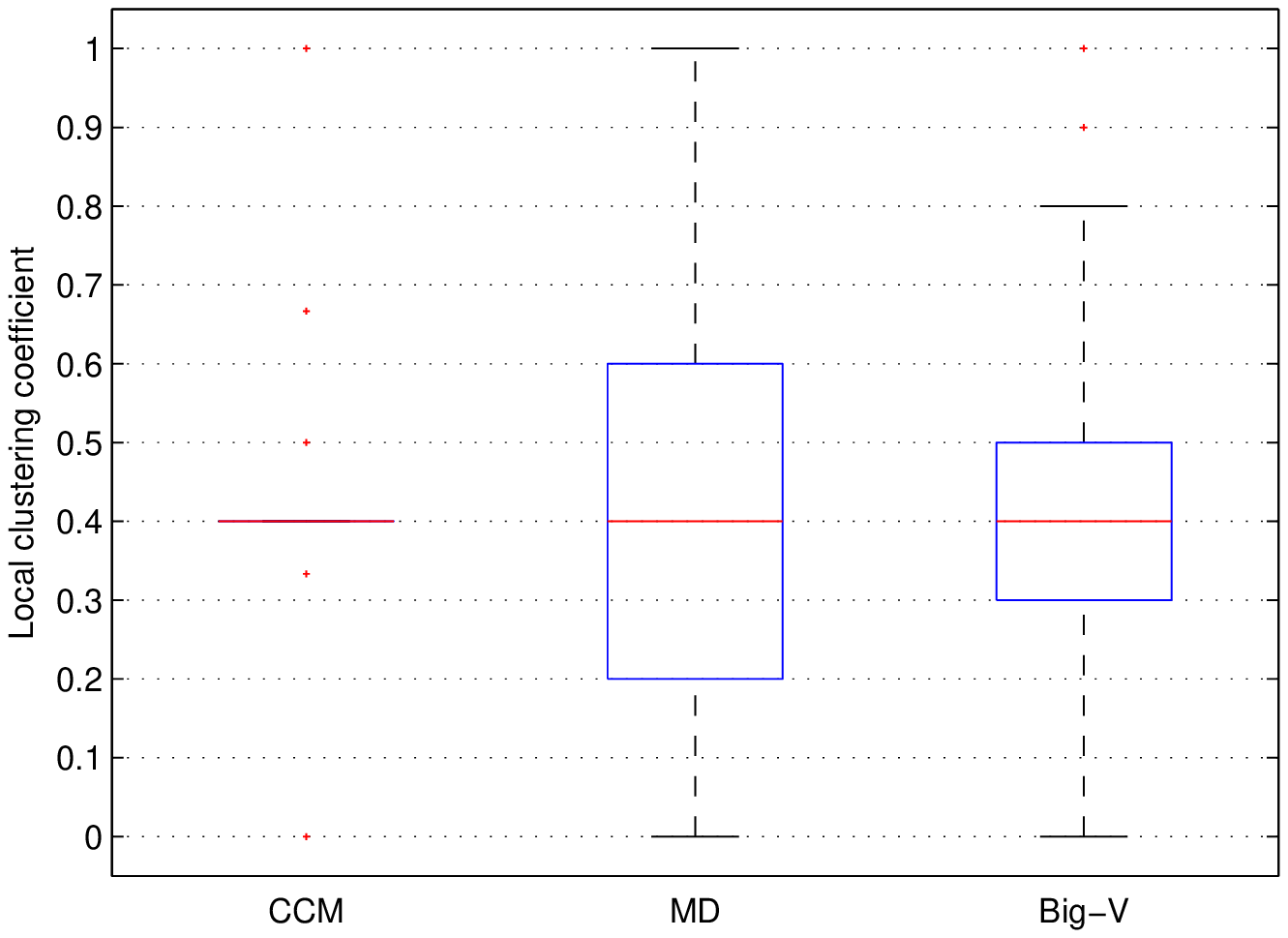}
  \caption*{$\phi =  0.4$}
\endminipage\hfill
\minipage{0.32\textwidth}%
  \includegraphics[width=\linewidth]{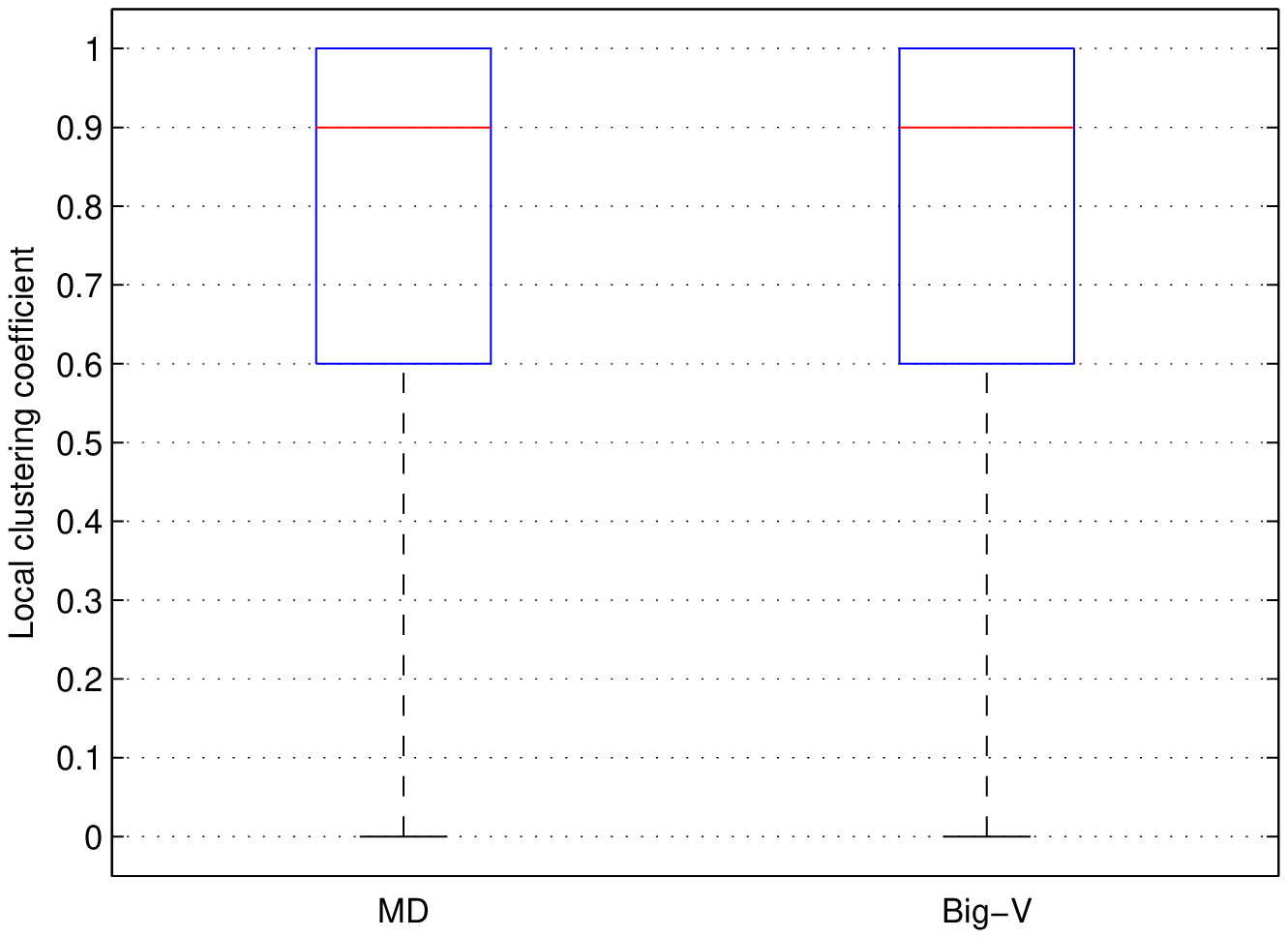}
  \caption*{$\phi =  0.8$}
\endminipage
\caption{Distribution of local clustering. Boxplots of local clustering measured from 20 homogeneous networks, $N=1998$, $\langle k \rangle = 5$. The size of the network was chosen to be divisible by $6$ due to the MD algorithm starting with disjointed fully-connected hexagons. Local clustering is a measure of interconnectivity between neighbours
of a given node. CCM shows an extremely tight distribution of clustering for $\phi=0.4$, as we would expect given that each node is allocated the same number and type of structure. MD provides the largest variance in local clustering.}
\label{fig:boxlc}
\end{figure}

\begin{figure}
\minipage{0.32\textwidth}
  \includegraphics[width=\linewidth]{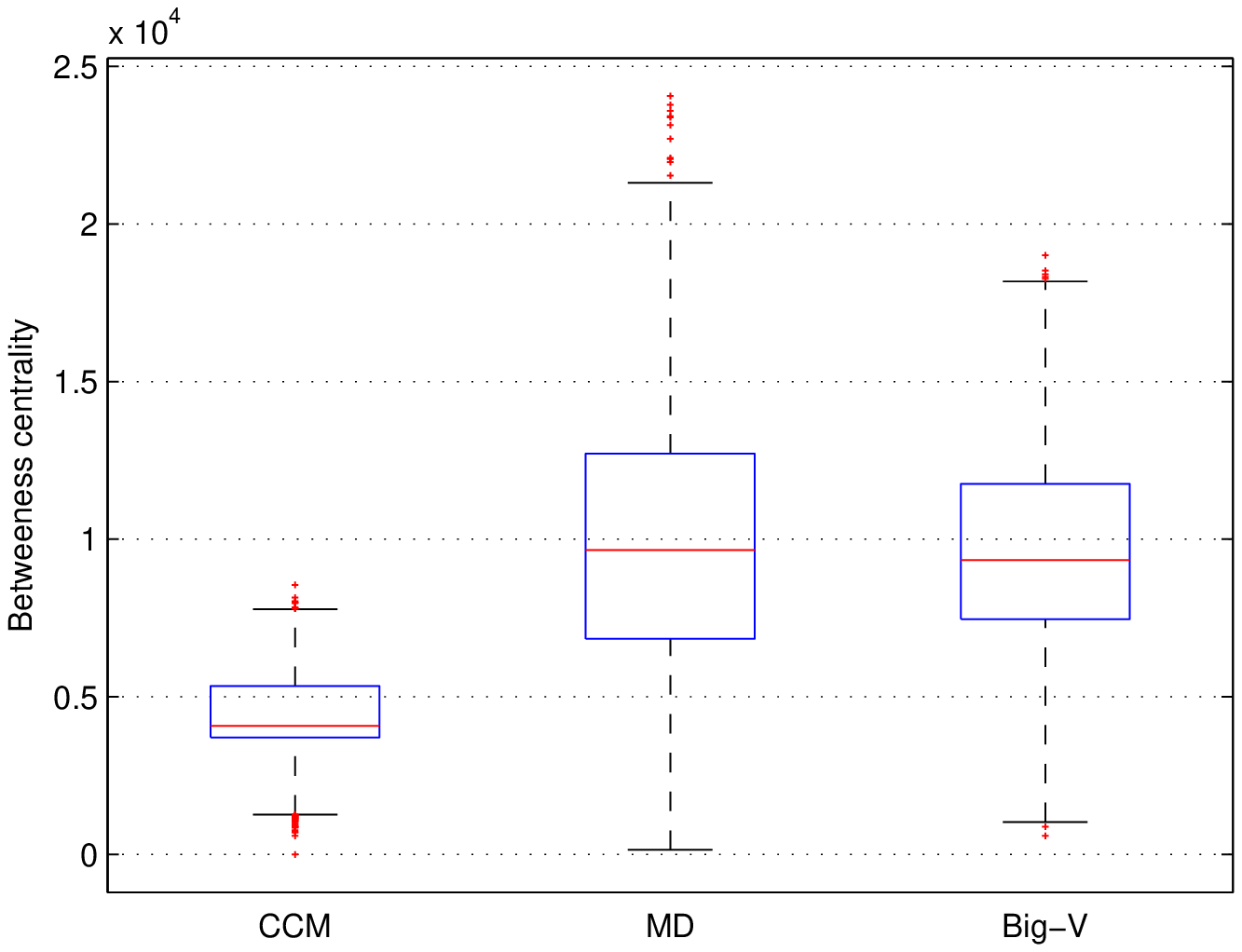}
  \caption*{$\phi =  0.2$}
\endminipage\hfill
\minipage{0.32\textwidth}
  \includegraphics[width=\linewidth]{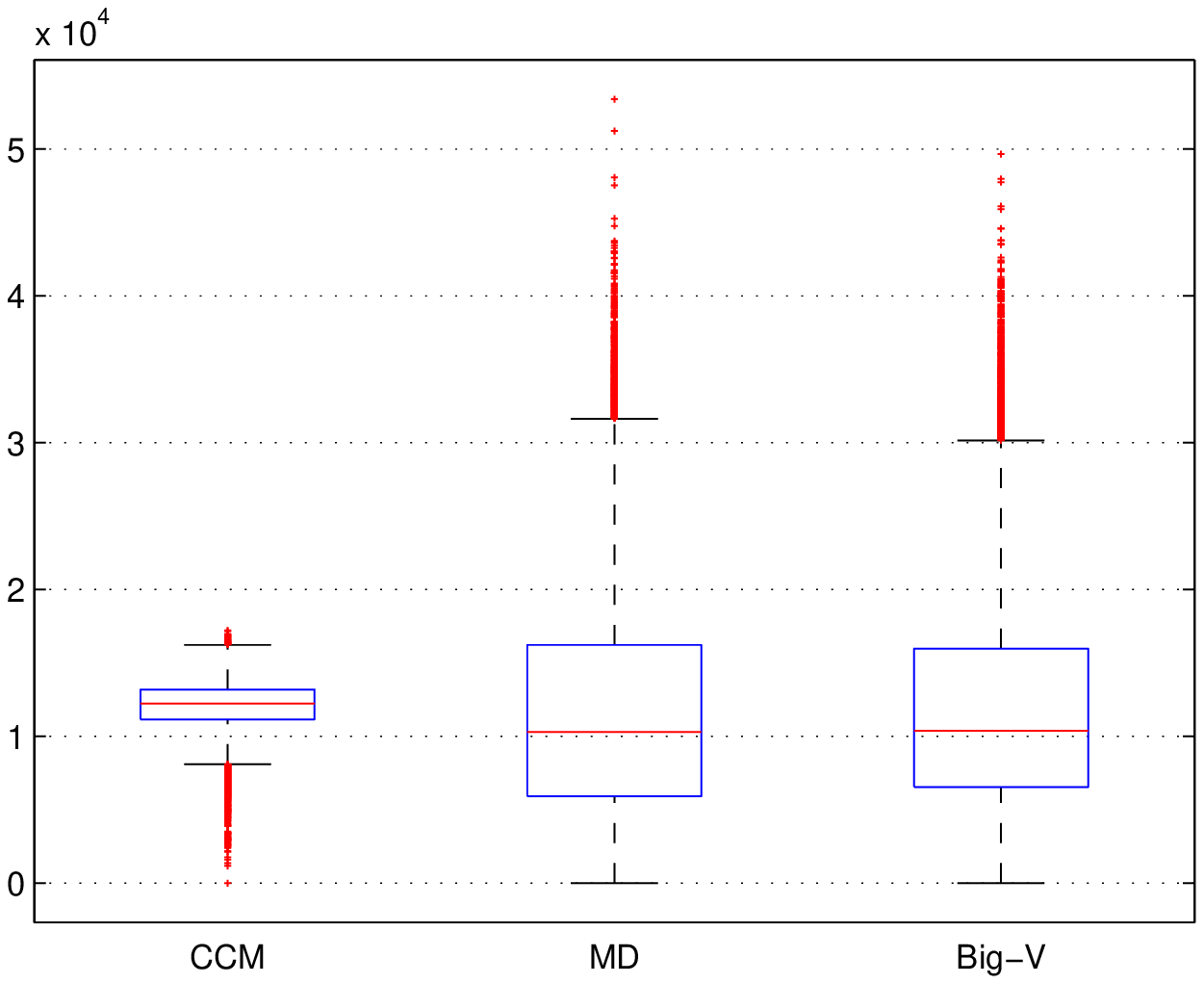}
  \caption*{$\phi =  0.4$}
\endminipage\hfill
\minipage{0.32\textwidth}%
  \includegraphics[width=\linewidth]{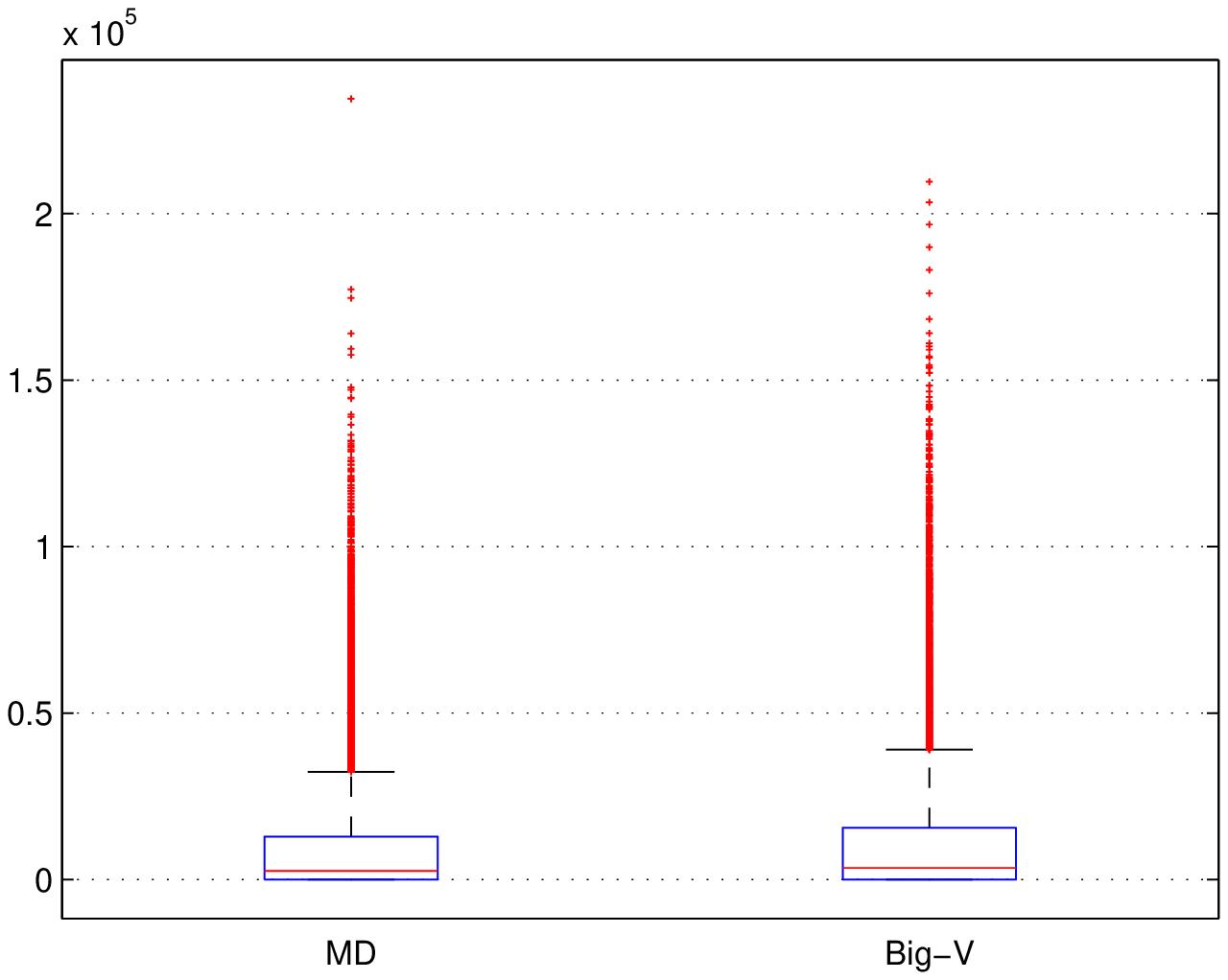}
  \caption*{$\phi =  0.8$}
\endminipage
\caption{Distribution of betweenness centrality. Box-plots of nodal betweenness centrality measured from 20 homogeneous networks, $N=1998$, $\langle k \rangle = 5$. Betweenness centrality ranks nodes on
how often they appear in paths between other nodes. As clustering is tuned higher, the CCM and Big-V rewiring algorithms isolate fully-connected clustered components of $\langle k \rangle + 1$ nodes away from the GCC. At this level of clustering highly-connected sets of nodes are still weakly attached to the GCC, yielding the large number of outliers observed in the plot, and hence, a high spread of betweenness centrality values.}
\label{fig:boxbc}
\end{figure}

\begin{figure}
\begin{center}
\includegraphics[scale=0.5]{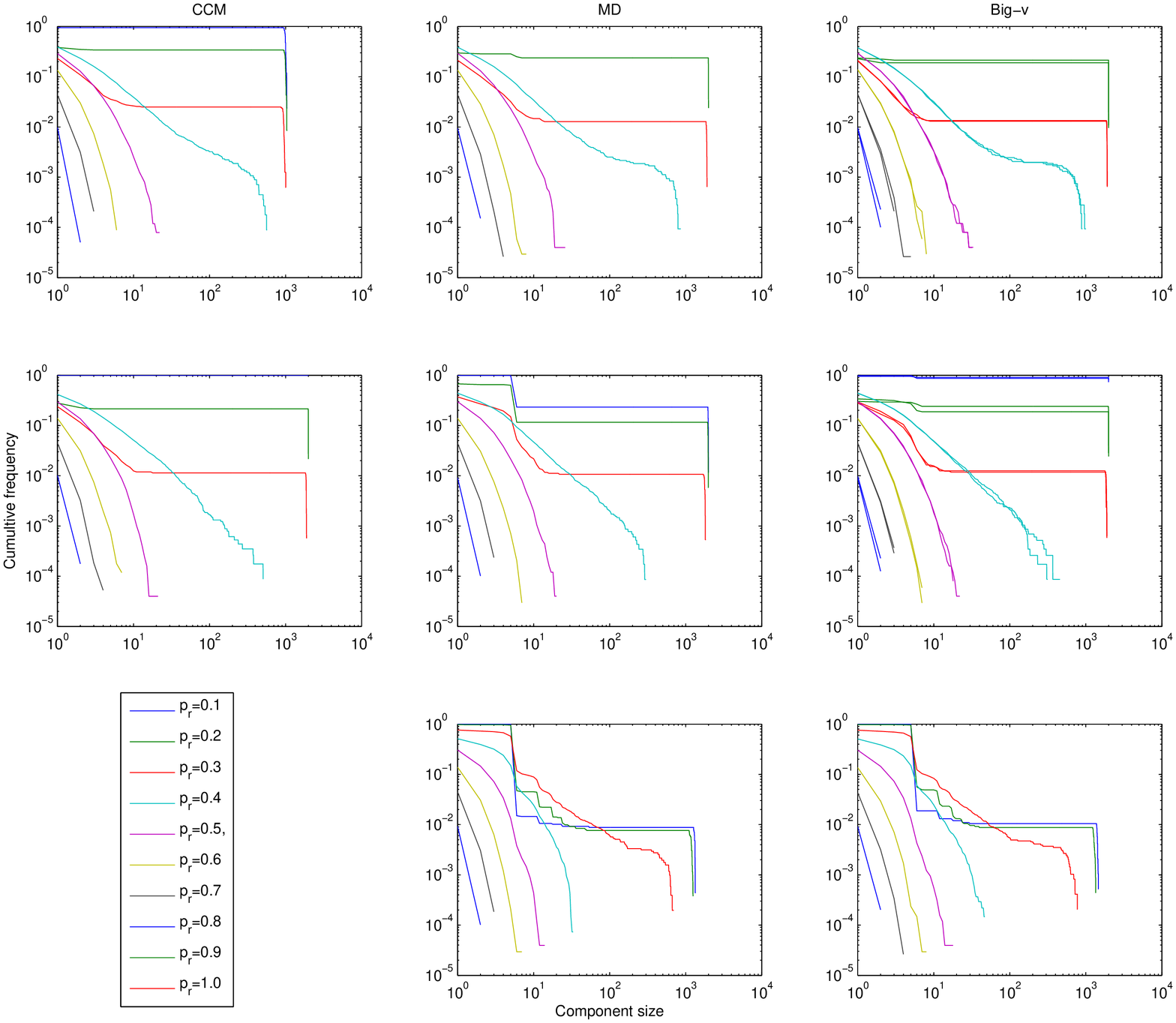}
\captionof{figure}{Edge percolation plots. Frequency of component sizes as edges are removed from the network with probability $p_r$. Results 
are taken from homogeneous networks with $\langle k \rangle = 5$ and $N = 1998$. CCM networks are not represented when $\phi=0.8$. Each line represents
a different value of $p_r$, varying from $0.1$ to $0.8$ in increments of $0.1$ . In each figure, $p_r$ increases in a clockwise direction.}
\label{fig:perc}
\end{center} 	
\end{figure}

\begin{enumerate}
%%%%%%%%%%%%%%%%%%%%%%%%%%%%%%%%
\item\noindent{\textbf{Overall feature and structure of the network:}}
%%%%%%%%%%%%%%%%%%%%%%%%%%%%%%%%
{\em Gephi} \cite{ICWSM09154} was used to visualize sample networks generated by the proposed algorithms, see Fig.~\ref{fig:gephi1}. In these figures nodes are colour coded
according to their degree of clustering, with un-clustered nodes coloured with nuances closer to the red end of the spectrum, and more highly clustered nodes coloured with
shades closer to the blue end of the spectrum. The figure clearly illustrates that the CCM algorithm gives rise to networks with an extremely homogeneous structure, whilst the 
rewiring algorithms (i.e. Big-V and MD) construct networks with more heterogeneity in clustering at node level. It is also evident that this difference translates 
into a more modular structure for the rewired networks. The CCM networks stand out as being structurally different from the networks generated by the other algorithms; 
as well as being homogeneous in degree, they are also homogeneous in structure.
%%%%%%%%%%%%%%%%%%%%%%%%%%%%%%%%
\item\noindent{\textbf{Distribution of clustering and centrality:}}
%%%%%%%%%%%%%%%%%%%%%%%%%%%%%%%%
The almost homogeneous distribution of the local clustering (see Fig.~\ref{fig:boxlc}) and betweenness centrality (see Fig.~\ref{fig:boxbc}) of the CCM networks is expected since by 
construction every node has the same local structure. For $\phi=0.2$ we know that each node has a quintuple of stubs with probability $p_1$ or a complete square and a triangle 
with probability $p_2=1-p_1$. When $\phi=0.4$ every node is a member of one triangle and one complete square.

The Big-V algorithm introduces clustering in more heterogeneous manner, with half of the nodes having clustering in the range $0.3 \leq \phi \leq 0.5$. The MD algorithm
provides the largest spread of clustering with half of the nodes having clustering in the range $0.2 \leq \phi \leq 0.6$. The box-plot (Fig.~\ref{fig:boxlc}) of local clustering shows the tendency 
of the MD algorithm to leave motifs unchanged. These complete motifs must be compensated with other parts of the network being decomposed into a much more random graph-type 
structure. The MD algorithm relies on random edge swapping to decompose the network into a more random structure. This will tend to result in at least a few leftover fully-connected motifs 
which are only destroyed at very low levels of clustering. Hence, when the frequency of such motifs is still large but the overall, desired, clustering is moderate, the connected parts of the network have to be left weakly clustered. 

The plot of betweenness centrality (Fig.~\ref{fig:boxbc}) illustrates this description in a more subtle way. Nodes embedded in a motif will have a low betweenness centrality whilst those
that act as bridges between structures and the rest of the network will be more highly ranked. We observe that the betweenness centrality plot shows a slightly higher spread
for MD networks. The removal of nodes (with a high betweenness centrality rank) is more likely to have a bigger impact on dynamics flowing on the network constructed by the MD algorithm. 
%%%%%%%%%%%%%%%%%%%%%%%%%%%%%%%
\item\noindent{\textbf{Connected component analysis:}}
%%%%%%%%%%%%%%%%%%%%%%%%%%%%%%%%
For a well connected network with low clustering, low values of $p_r$ leave the macroscopic structure of the networks unchanged (see Fig.~\ref{fig:perc})
i.e. the entire network is contained within a single GCC. Networks with low clustering are resilient to the removal of a relatively small number of edges. 
This behaviour has been previously noted \cite{green2010large}. 

Reading row-wise across Fig.~\ref{fig:perc}, for $\phi=0.4$, the GCCs of CCM networks are the most robust to edges being removed. We know the networks to be
highly homogeneous such that no particular edge is structurally more important than another. MD networks are extremely sensitive to edges being removed; 
$p_r=1/40$ has a pronounced impact revealing a strong dependency on components of size 6 or fewer. The MD algorithm tends to preserve some complete motifs,
as well as to leave some motifs weakly connected to the GCC at this level of clustering. The Big-V algorithm generates networks with a relatively
well connected GCC but still exhibits a mild sensitivity to the removal of edges. 

%%%%%%%%%%%%%%%%%%%%%%%%%%%%%%%%%%%%%%%%%%%%%%%%%%%%%%%%%%%%%%%%%%%%%%%%%%%%%%%%%%%%%%%%%%
\item\noindent{\textbf{Motif statistics for all network types:}}
%%%%%%%%%%%%%%%%%%%%%%%%%%%%%%%%%%%%%%%%%%%%%%%%%%%%%%%%%%%%%%%%%%%%%%%%%%%%%%%%%%%%%%%%%

\begin{table}
\begin{center}
    \begin{tabular}{l|l|l|l|l}
    \hline
           ~                   & $\phi^1_4$ & $\phi^2_4$ $\boxempty$ & $\phi^3_4$ $\boxslash$  & $\phi^4_4$ $\boxtimes$ \\ \hline
     CCM,~~$\phi = 0.2$    & 0.0053     & 0.0007     & 0.0001      & 0.0045     \\ 
    Big-V, $\phi = 0.2$    & 0.0117     & 0.0004     & 0.0104      & 0.0009    \\ 
    MD,~~~ $\phi = 0.2$ & 0.0239     & 0.0057     & 0.0149      & 0.0034     \\  \hline  
      CCM, ~$\phi = 0.4$  & 0.0169     & 0.0003     & 0.0002      & 0.0164    \\  
    Big-V, $\phi = 0.4$    & 0.0570     & 0.0010     & 0.0400      & 0.0160    \\
       MD, ~~~$\phi = 0.4$ & 0.0731     & 0.0083     & 0.0444      & 0.0204    \\ \hline
    Big-V, $\phi = 0.8$    & 0.3150     & 0.0013     & 0.0900      & 0.2237    \\
       MD, ~~~$\phi = 0.8$ & 0.3405     & 0.0044     & 0.1062      & 0.2299    \\
    \end{tabular}
\end{center}
\caption{For each level of clustering the table has been sorted in ascending $\phi$ then ascending $\phi^1_4$, where $\phi^1_4$ gives the proportion of all closed quadruples. The above table is computed using unique counts.}
\label{table:tab1}
\end{table}

Table \ref{table:tab1} shows that third-order clustering conveys little information about order-four motifs. 
%Whilst, for example, fully connected order-four structures cannot exist if there are no triangles - and conversely, we would expect very few squares without diagonals if there were many triangles - in the intermediate regions these motif prevalences are not strongly determined by the clustering coefficient.
The proportion of closed quadruples to all connected structures of 4 nodes increases with clustering, and for high levels of clustering there is a strong dependence on complete squares. However, the algorithms' lack of control of order-four structure is apparent at moderate levels of clustering ($\phi =  0.2,~0.4$) where there is no consistent presence of closed order-four structures across networks of equal clustering. The difference in $\phi^1_4$ is due to triangles which do not share an edge; indeed these triangles are not measured by this metric. The distribution of triangles is important at higher levels of clustering where they often share edges or overlap to form order-four structures. 

Reading column-wise down Fig.~\ref{fig:motifdist}, we see a more particular dependence on squares with diagonals as clustering increases. For $\phi=0.4$ there is the greatest heterogeneity in the distribution of diagonal squares. When $\phi = 0.8$ we observe that diagonals can only appear in certain combinations about a node, following an almost tri-modal distribution. Again reading column-wise down, for complete squares we see a general trend of increasing complete square prevalence with increased clustering. Nodes may have a count of ten complete squares associated with them when they are members of a complete, and isolated, six-node structure. At all levels of clustering the probability of finding an empty square associated with a node is rare. This is because, despite being a structure of order-four, squares do not contribute to clustering. 

Finally, Fig.~\ref{fig:motifdist} also reveals that networks generated by the Big-V algorithm contain empty square motifs with very low frequency. The algorithm searches for unclosed triples contained within strings of five nodes and closes them. Only motifs that may be constructed out of triangles can be expected in Big-V networks in any significant quantity. The MD algorithm also generates few empty square motifs and the CCM algorithm will only include them by specification. 
\end{enumerate}

%%%%%%%%%%%%%%%%%%%%%%%%%%%%%%%%%%%%%%%%%%%%%%%%%%%%%%%%%%%%%%%%%%%%%%%%%%%%%%%%%%%%%%%%%
\subsection{Dynamics on the networks: evaluation and comparison}
\begin{figure}
\begin{center}
\begin{tabular}{cc}
 \includegraphics[scale=0.45]{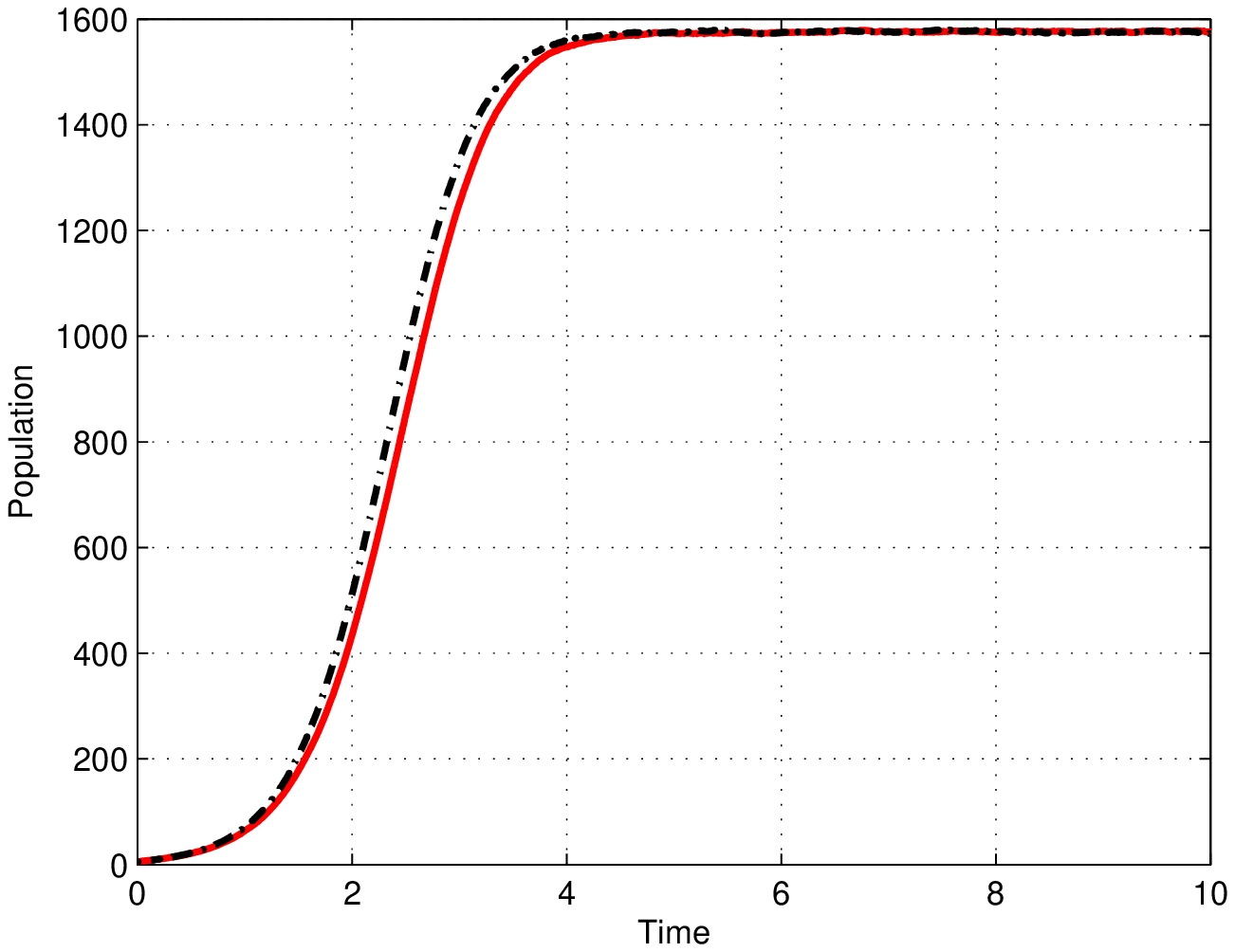}  & 
 \includegraphics[scale=0.45]{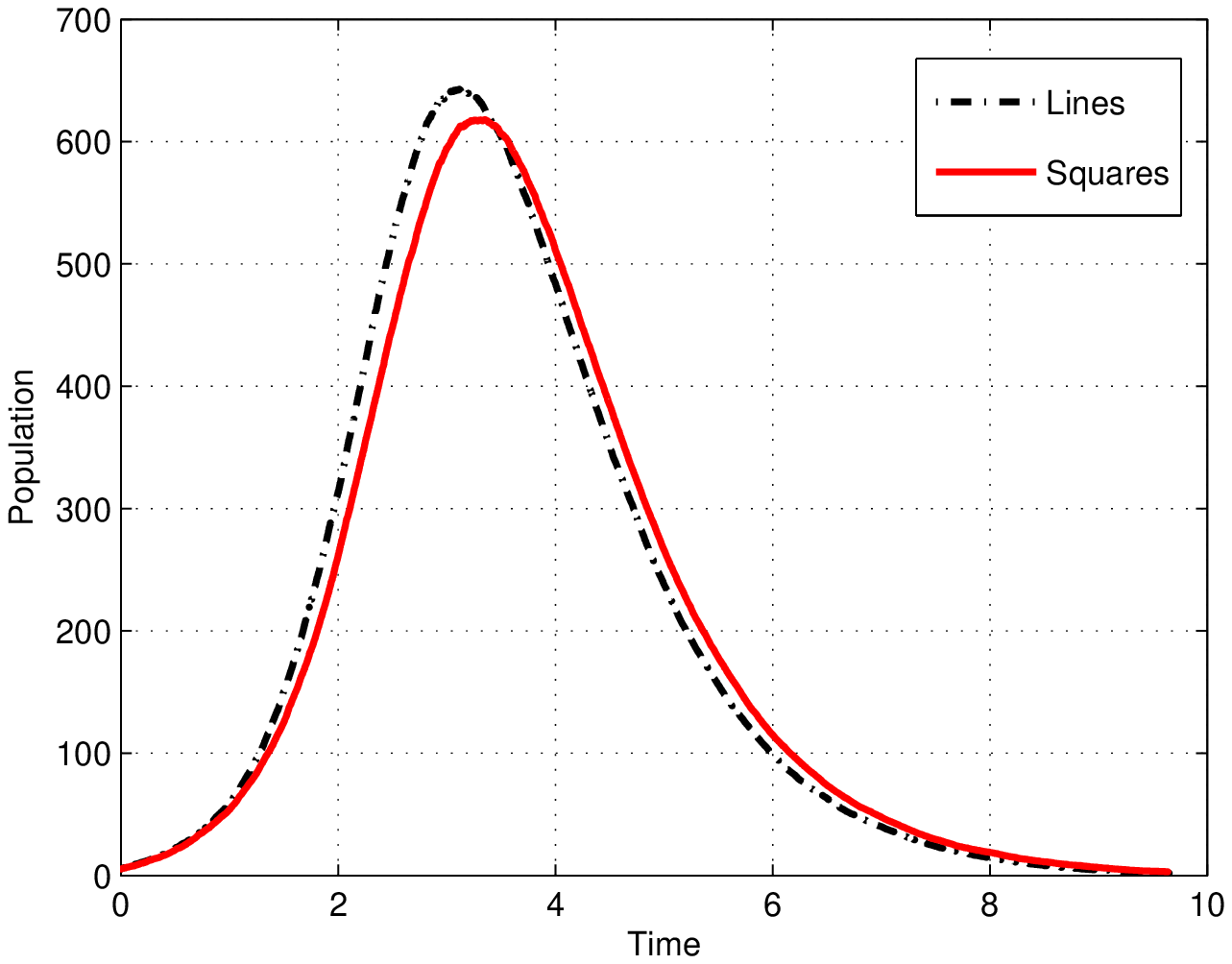}  \\
	SIS & SIR\\
\end{tabular}
\caption{Random vs square comparison. 20 homogeneous networks were generated with: $N = 1998$, $\langle k \rangle = 5$ and $\phi=0.0018$. The plots correspond to averaging Gillespie simulations on each of the networks with parameters $\tau = \gamma  = 1$, and $5$ initially infectious nodes. The networks marked `square' were constructed by allowing two squares to be formed out of a nodes 5 stubs, compared against a random network. The plots show comparisons between the prevalence of infection for $SIS$ and $SIR$ dynamics.}
\label{fig:linesvsquares}
\end{center}
\end{figure}

%%%%%%%%%%%%%%%%%%%%%%%%%%%%%%%%%%%%%%%%%%%%%%%%%%%%%%%%%%%%%%%%%%%%%%%%%%%%%%%%%%%%%%%%%
The effect of higher-order structure on epidemics is not obvious. For triangles it is observed that when an initially infected individual
infects a second, the two infected nodes then compete for the same remaining susceptible. For empty squares and longer loops the effect is similar but less dramatic. Fig.~\ref{fig:linesvsquares}
shows that the initial epidemic spread is slower for networks which exhibit loops. 
By opening a closed motif whilst preserving degree, two new individuals must be added so the effect of competition is inversely proportional to the motif size. Connectivity within the motif may also negate the effect of competition. 

\begin{figure}
% \begin{center} 
\begin{tabular}{ccc}
 \includegraphics[scale=0.35]{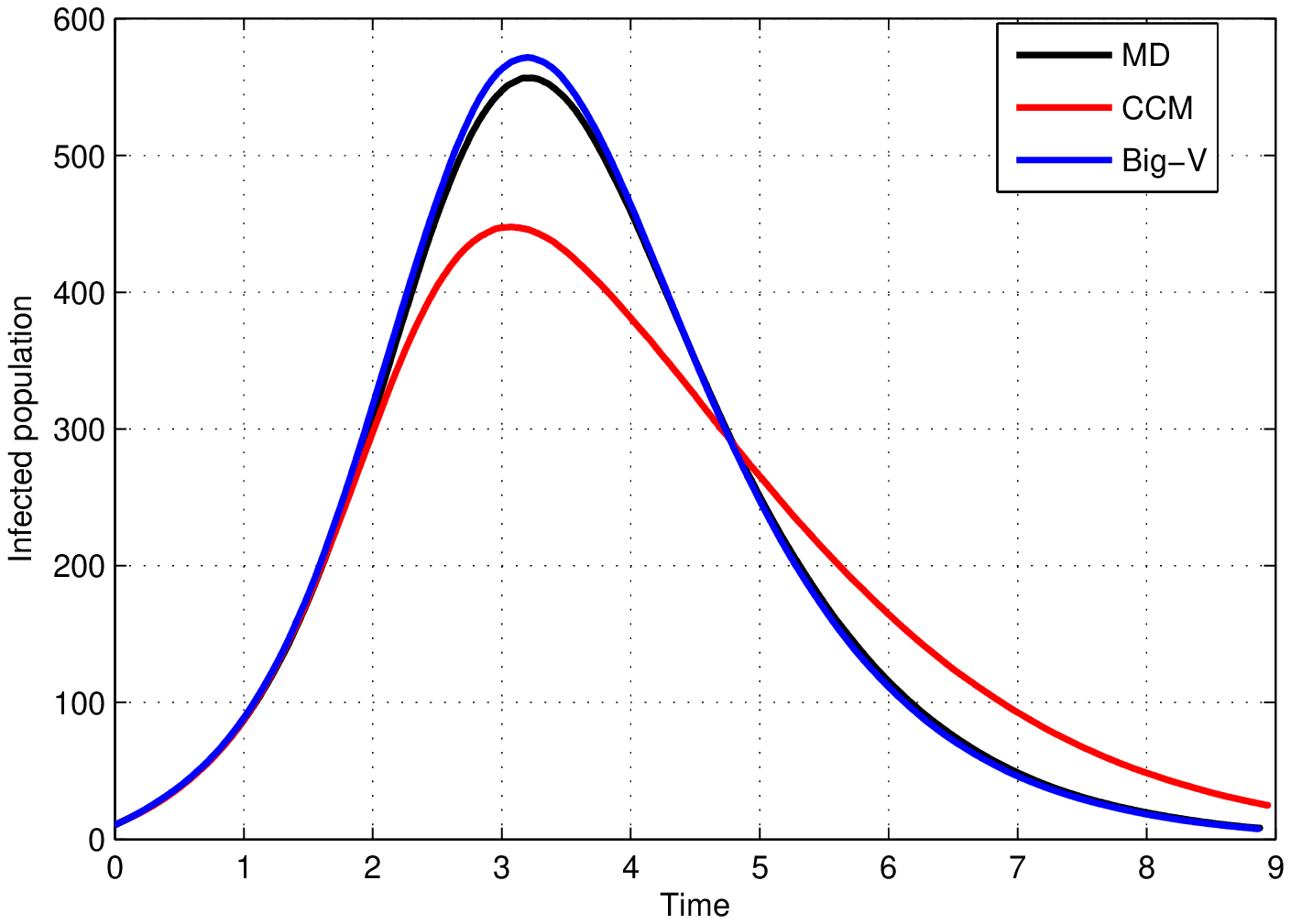} & 
 \includegraphics[scale=0.35]{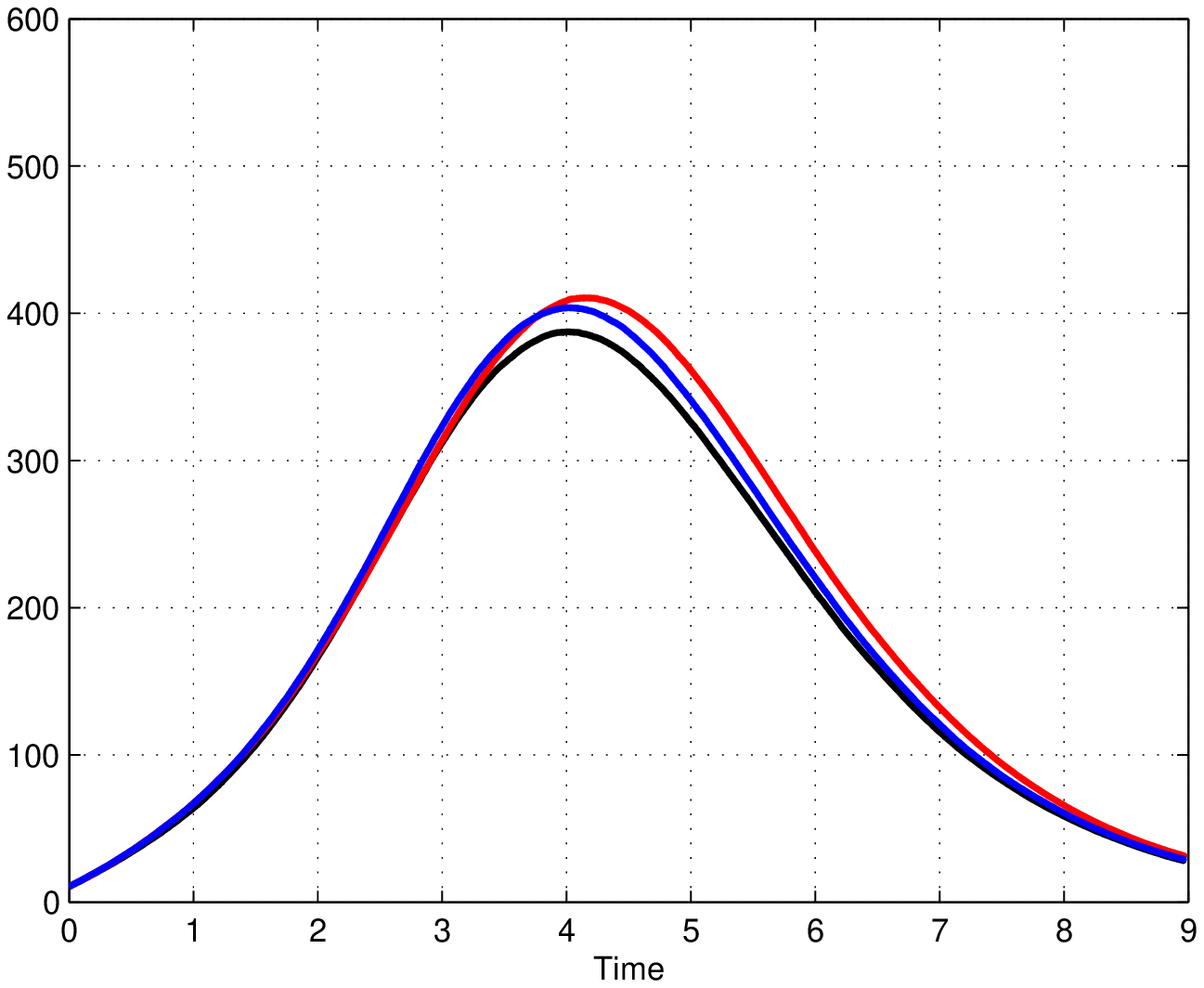} &  
 \includegraphics[scale=0.35]{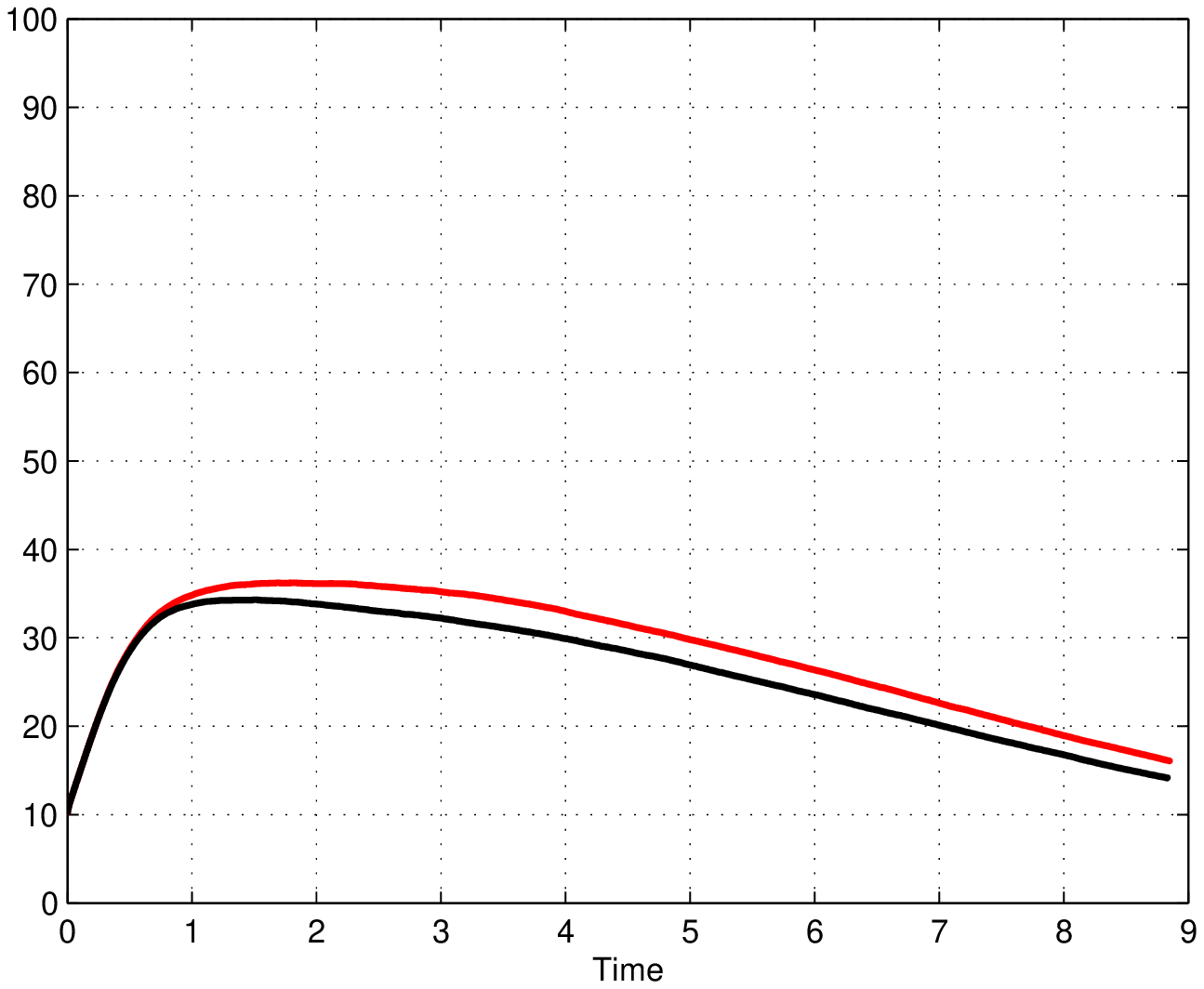} \\
 \includegraphics[scale=0.35]{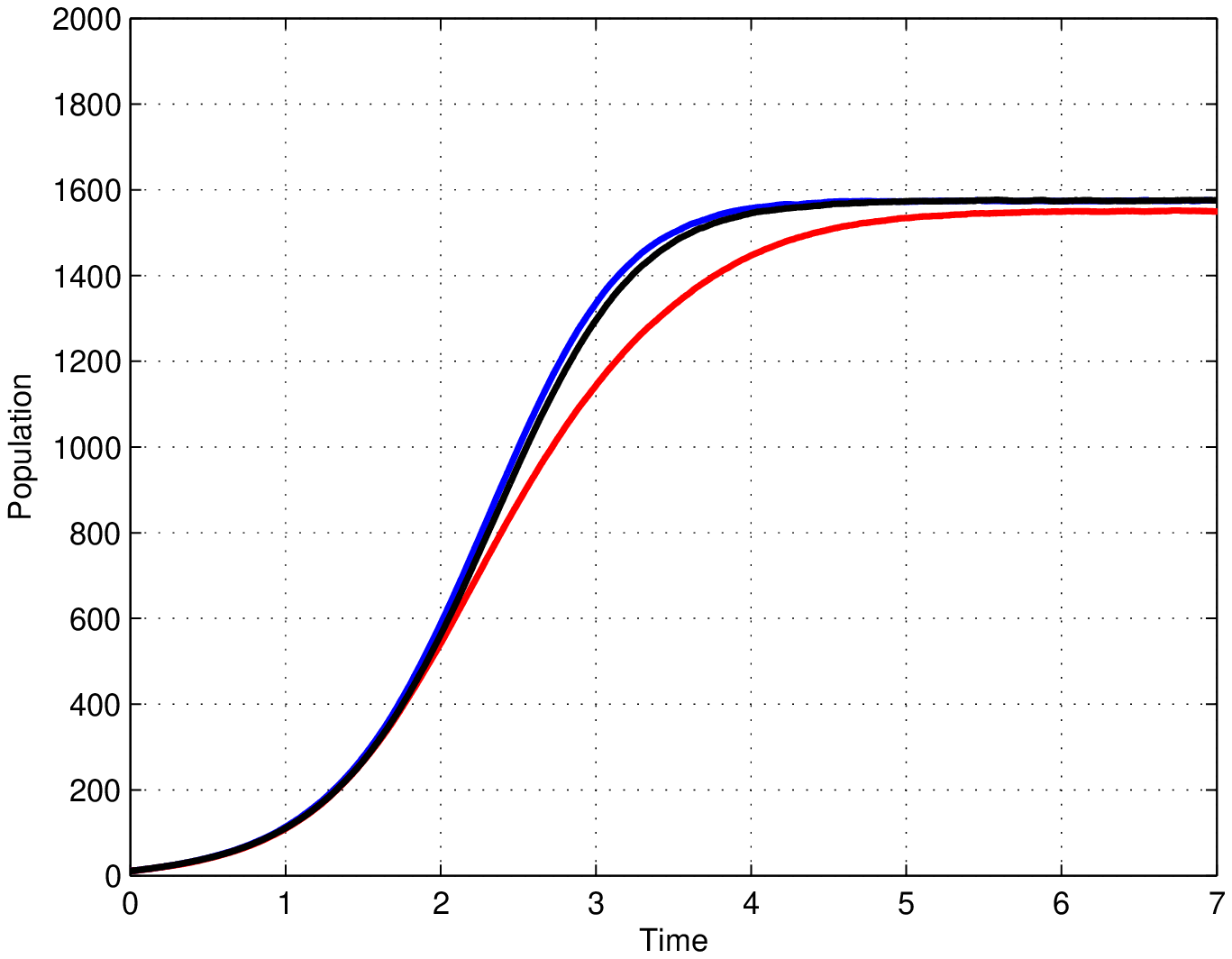} &
 \includegraphics[scale=0.35]{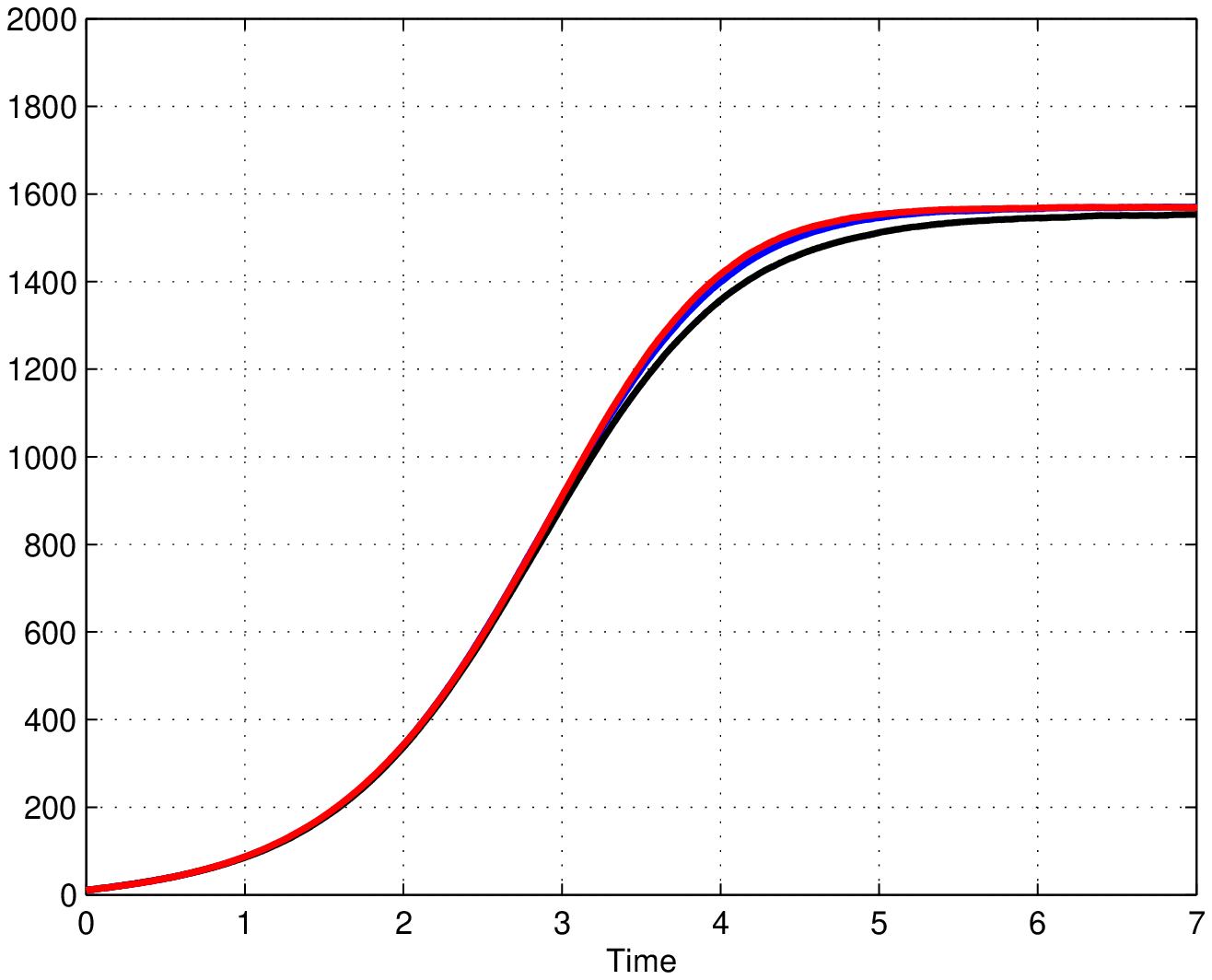} &
 \includegraphics[scale=0.35]{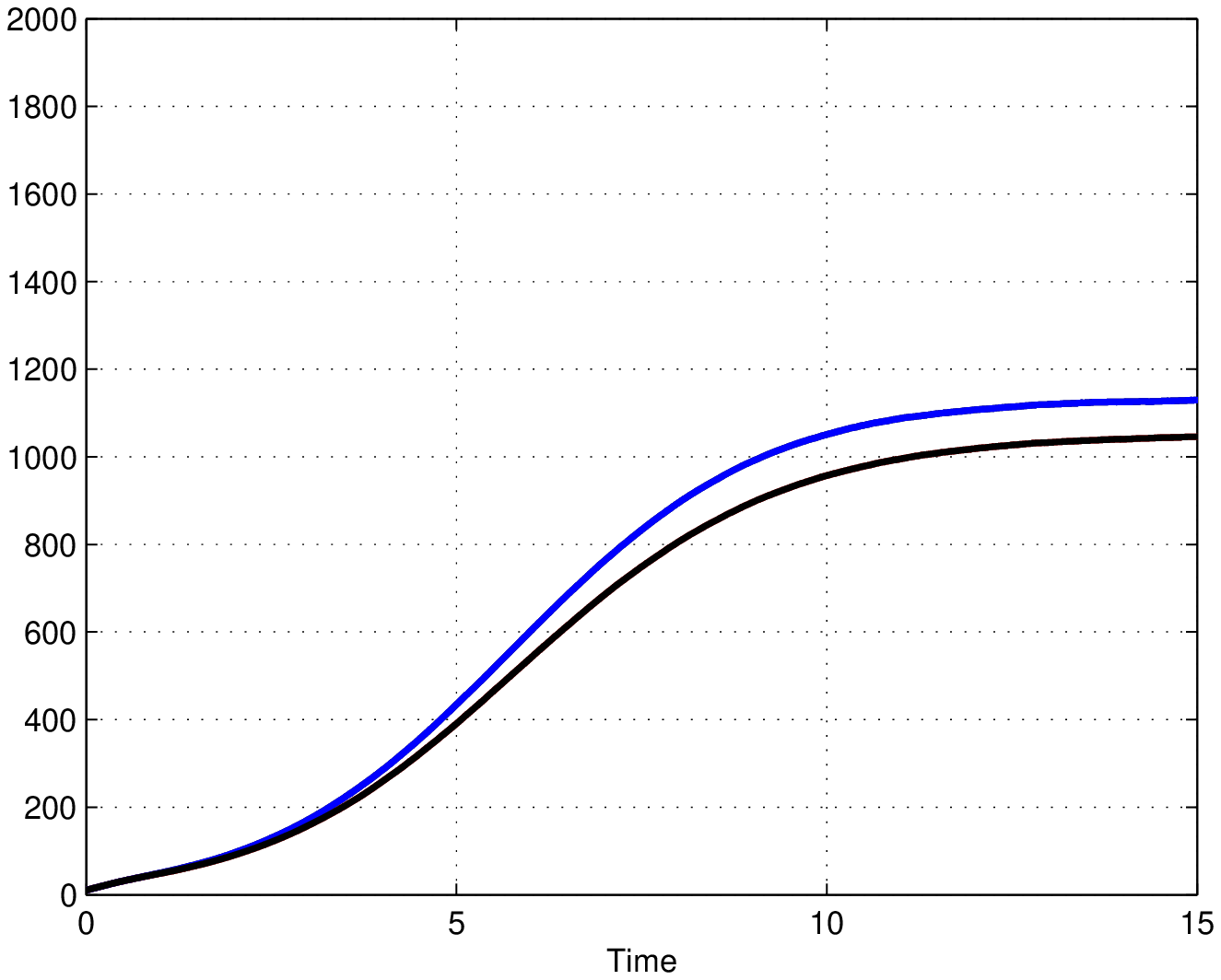} \\
 $\phi = 0.2$ &  $\phi = 0.4$ &  $\phi = 0.8$ \\
\end{tabular}
\caption{SIR and SIS dynamics. 20 homogeneous networks were generated with: $N = 1998$, and $\langle k \rangle = 5$, and the results show the average of $100$ Gillespie epidemics on each network realisation. The epidemics were run with parameters $\tau = \gamma  = 1$, and were seeded with 5 infectious nodes. The top and bottom rows show the prevalence levels for a $SIR$ and $SIS$ epidemics, respectively.}
% \end{center}
\label{fig:epidemic}

\end{figure}

When simulating epidemics on networks with $\phi=0.2$, the CCM networks show a slower spread of infection (Fig.~\ref{fig:epidemic}).  At this level of clustering 
the CCM algorithm breaks a quintuple of stubs into all lines with $p=1/2$, or a complete square corner and a triangle corner with $p=1/2$. Thus, the CCM networks exhibit
areas of high clustering in which the disease will spread more slowly than in areas of low clustering. At $\phi = 0.2$ the CCM networks exhibit a slower spread of infection
for both SIS and SIR epidemics.  Reading row-wise from left to right it is clear that higher levels of clustering slows the epidemic, see the difference between $\phi=0.2$ and $\phi=0.4$, with a less
dramatic effect for $SIS$ epidemics. Tuning clustering to an even higher level leads to the network breaking down into many disjointed components, such that connectivity within these is excellent.
This means that the initial spread could be very fast, but this is quickly curtailed by limited or no connectivity between the highly connected components.
 
\begin{figure}
\begin{tabular}{ccc}
 \includegraphics[scale=0.35]{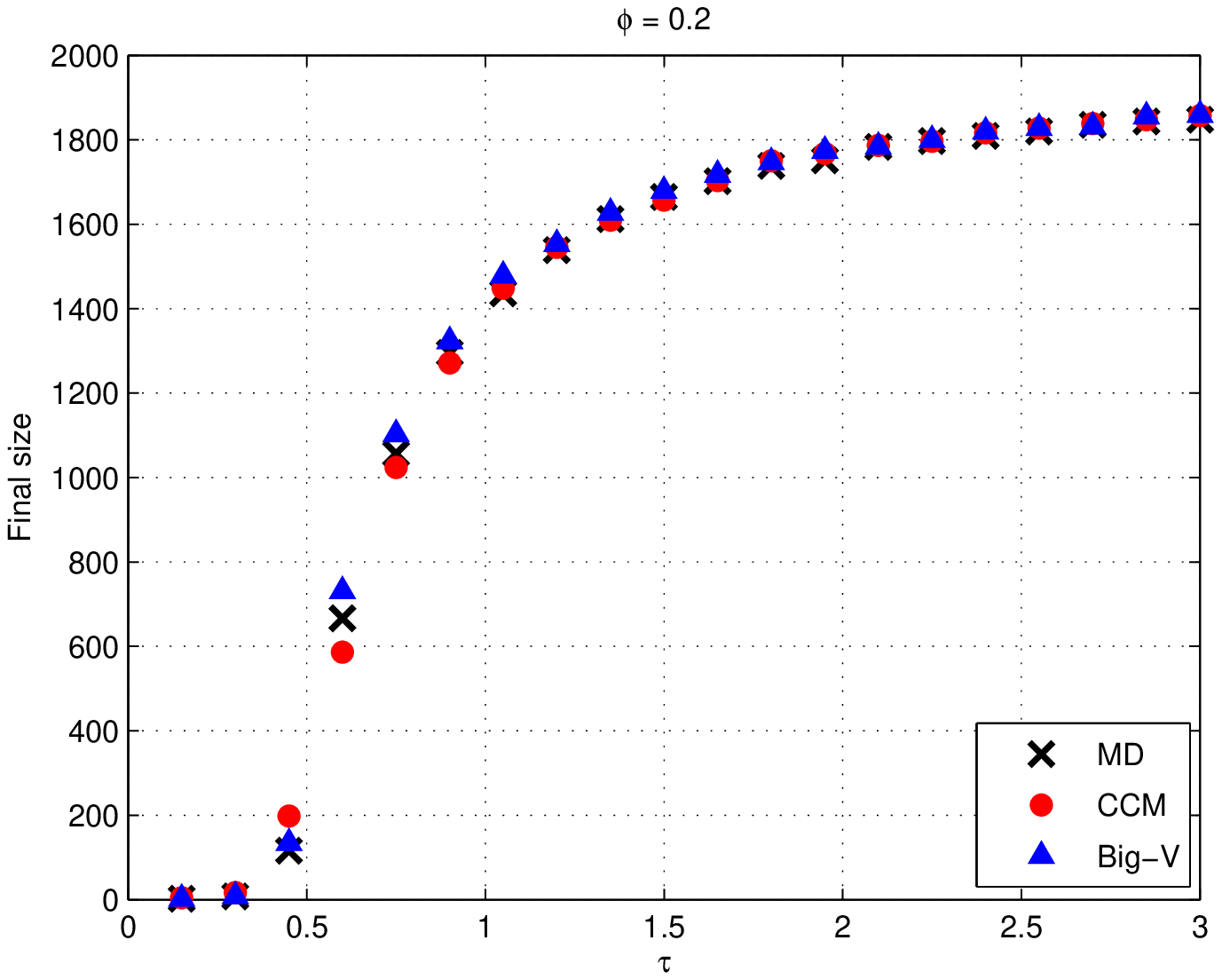} & 
 \includegraphics[scale=0.35]{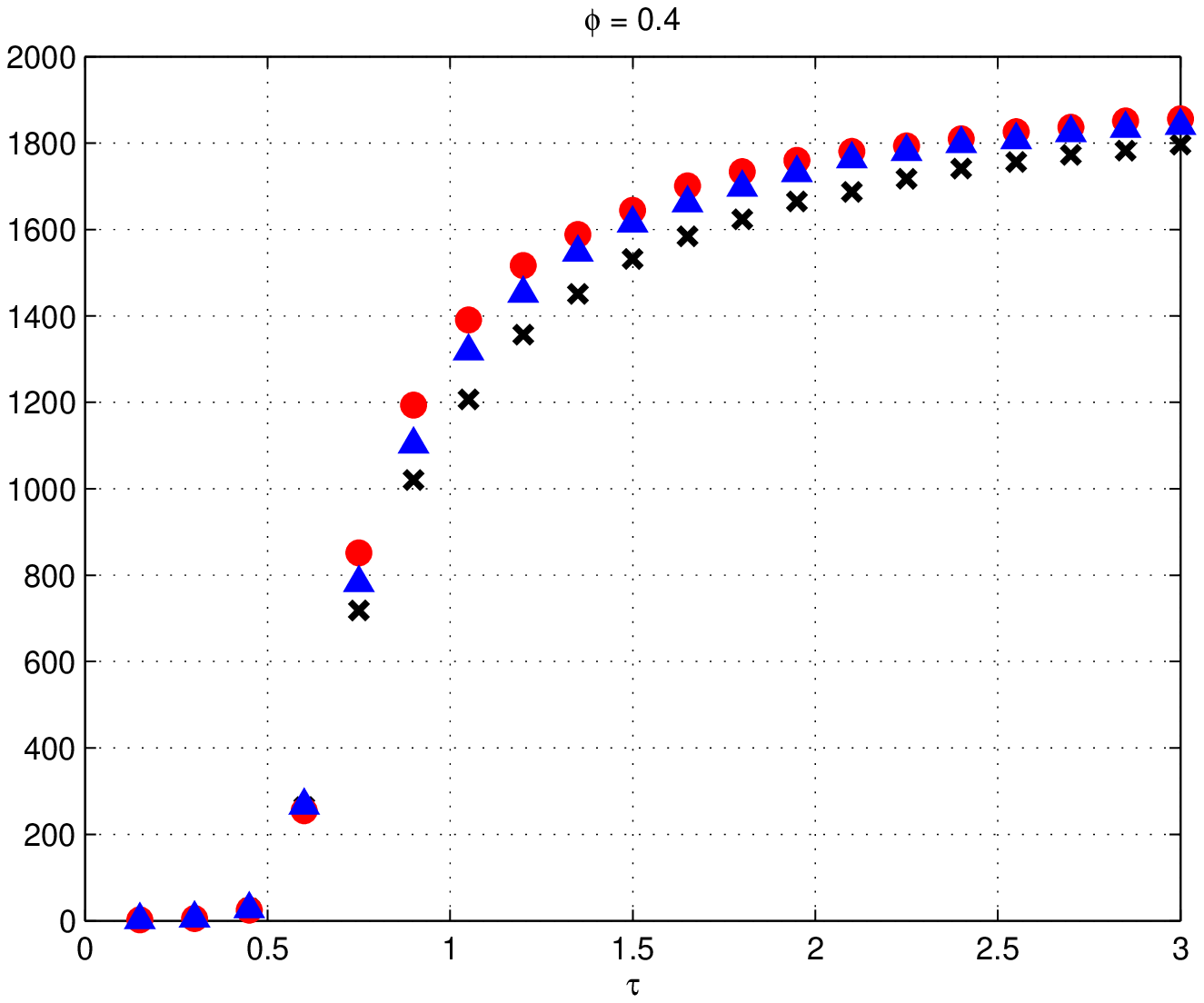} &  
 \includegraphics[scale=0.35]{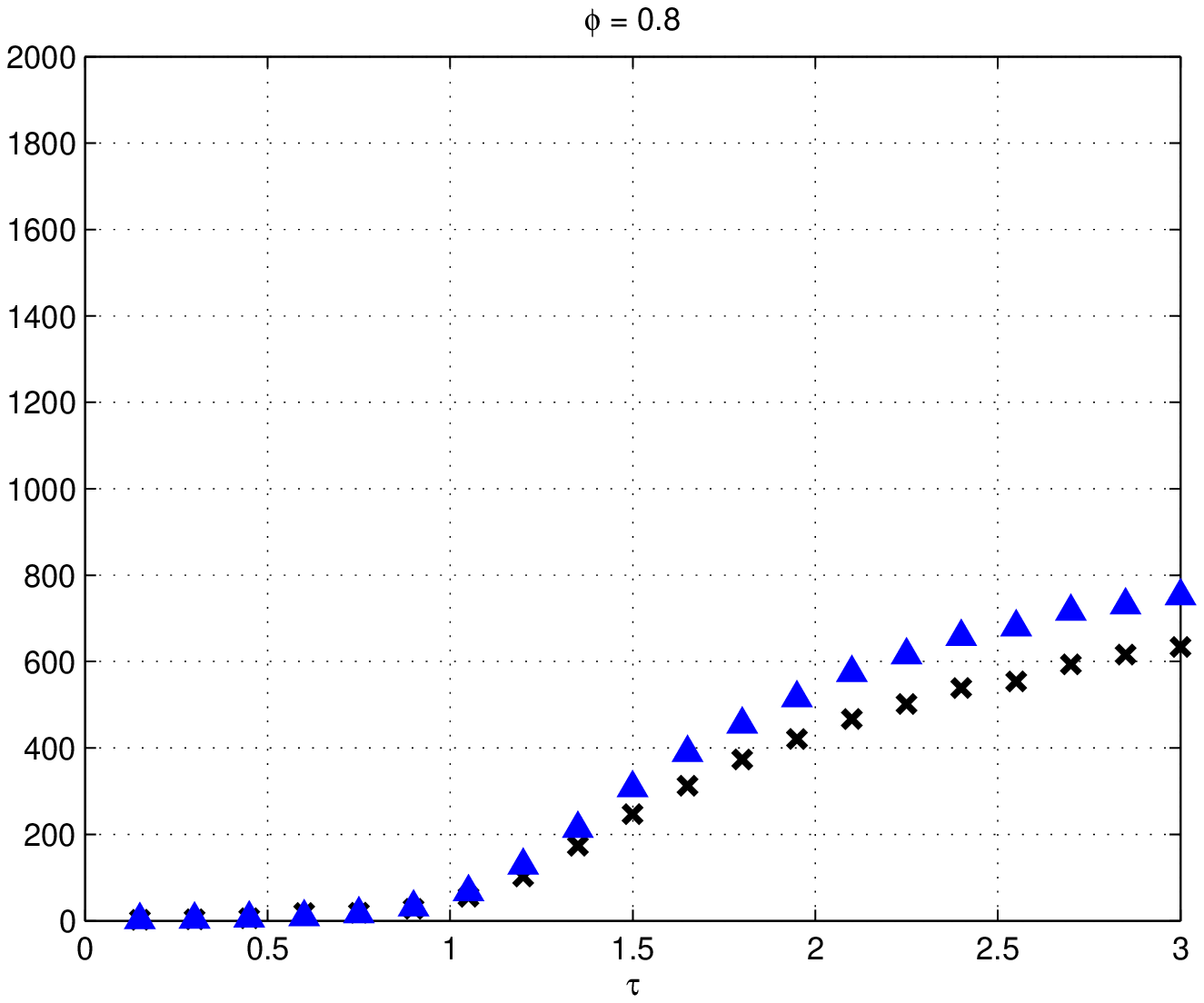} \\
 \includegraphics[scale=0.35]{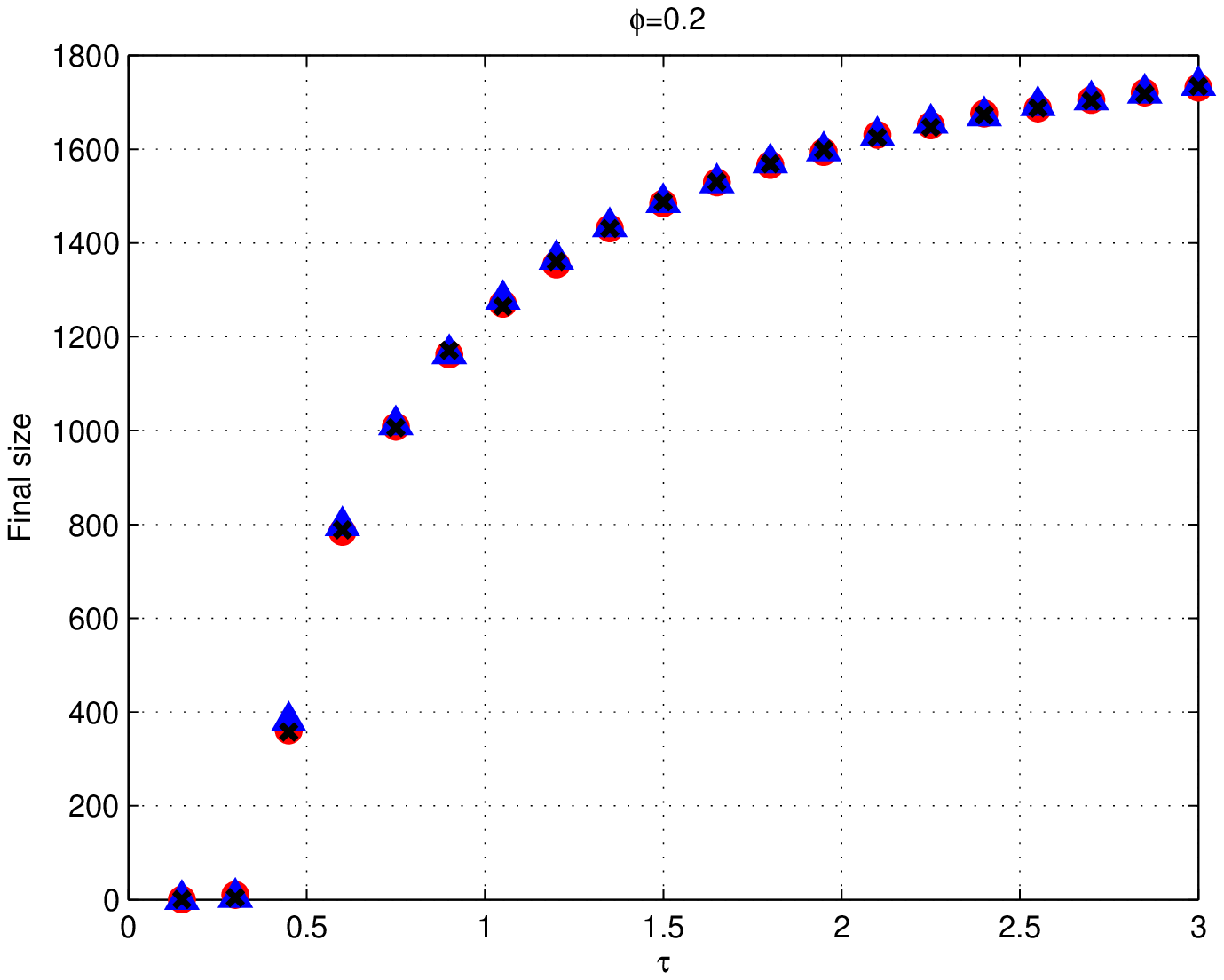} & 
 \includegraphics[scale=0.35]{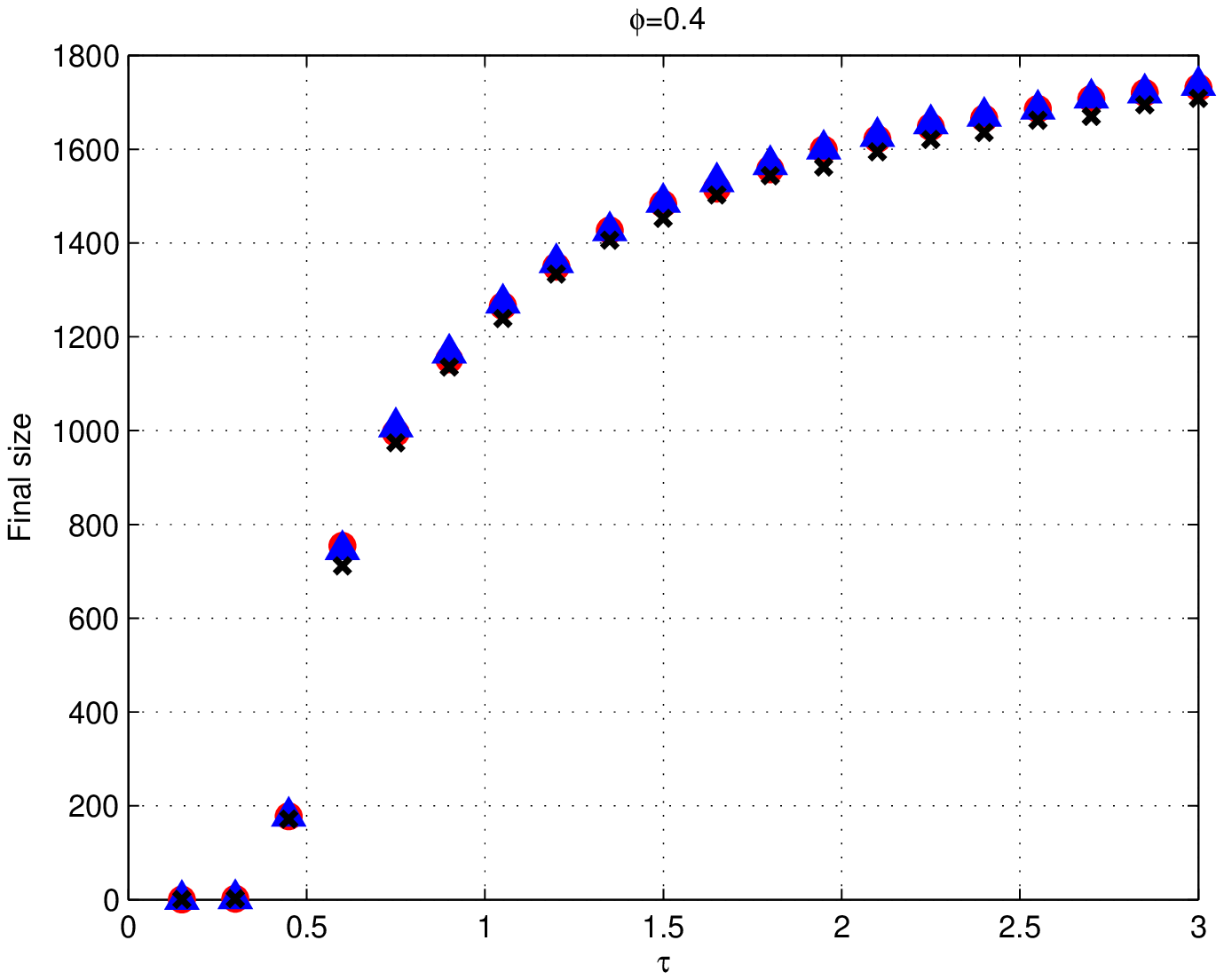} &  
 \includegraphics[scale=0.35]{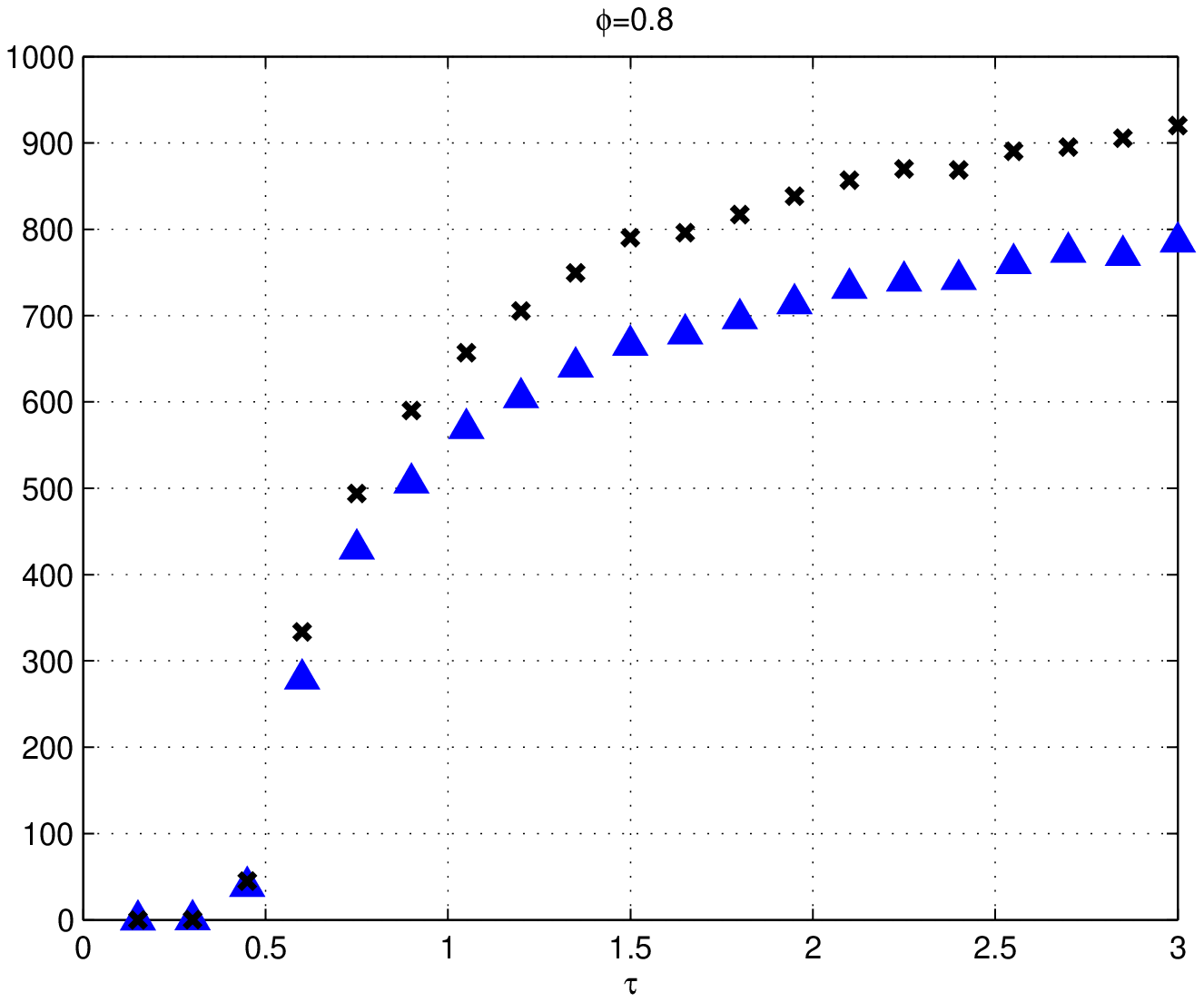} \\
  $\phi = 0.2$ &  $\phi = 0.4$ &  $\phi = 0.8$ \\
\end{tabular}
\caption{Plots of final epidemic size  (top row) and endemic equilibrium (bottom row) for values of $\tau$ increasing from $\tau = 0.1$ to $\tau=3$ in increments of $0.15$. Twenty networks were generated, ten Gillespie simulations were performed for each value of $\tau$.  The networks were homogeneous with $\langle k \rangle =  5$ and $N = 1998$.}
\label{fig:varytau}
\end{figure}

The rewiring algorithms tend to produce networks that contain clustered motifs that are poorly connected to the rest of the networks. 
Nodes with high betweenness centrality are important in SIR-type processes, which when recovered significantly hinder the propagation of the epidemic. 
All of the rewiring algorithms produce nodes with high betweenness centrality when compared to the CCM algorithm. It has previously been noted that the MD networks
are particularly dependent on isolated motifs; this is reflected in consistently smaller final epidemic size. The CCM networks have a more consistent connectivity throughout the network
yielding a greater final size (see Fig.~\ref{fig:varytau}). 

%%%%%%%%%%%%%%%%%%%%%%%%%%%%%%%%%%%%%%%%%%%%%%%%%%%%%%%%%%%%%%%%%%%%%%%%%%%%%%%%%%%%%%%%%%%%%%%%%%
\section{Discussion}
%%%%%%%%%%%%%%%%%%%%%%%%%%%%%%%%%%%%%%%%%%%%%%%%%%%%%%%%%%%%%%%%%%%%%%%%%%%%%%%%%%%%%%%%%%%%%%%%%%

The development of models that capture epidemic or other dynamics on networks is guided, to a great extent, by the structure of the network. Hence, models have initially sought to account for the impact of degree distribution, or heterogeneity in contact. This was closely followed by models capturing preferential mixing, where nodes of similar degrees can be 
either more likely (assortative mixing) or less likely (disassortative mixing) to be connected. The next stages of model development 
considered all the above, or at least accounting for their effect,  but looked at clustering, the propensity that neighbours of a node are likely to
be also neighbours of each other. In this area, progress is still being made and there are many new developments to follow. 

Many network models operate on and use synthetic networks designed to be able to control and tune properties such as degree distribution, mixing, clustering and so forth. 
However, as shown in this paper, controlling certain lower order (node, contact, neighbourhood) properties can and will
have an effect on higher-order structure and this can be significant and cannot be disregarded. 
For example, at high values of clustering generated based on the spatial algorithm \cite{read2003disease}, the
networks become more assortatively mixed. Such effects are to be expected since the network is a coherent structure which reacts to each perturbation, such as rewiring or other means of tuning properties.
In general, we expect that Big-V rewiring will be the most random way to introduce clustering without model-specific artefacts. However, this comes at what is occasionally prohibitive computational cost. CCM is computationally cheap, and analytically tractable, but is a long way from randomly introduced clustering. In this context, MD can be viewed as a computationally cheap and (given sufficient effort) analytically tractable alternative to Big-V that produces very similar network phenomenology.

In this study, we highlighted that synthetic algorithms that generate networks with tunable clustering do lead to different higher-order structures, 
such that networks with the same degree distribution and level of clustering can yield different dynamics on the networks. 
In order to evaluate differences in higher-order structures we have extended the concept of clustering and proposed some measures to evaluate 
and quantify the frequency of structures composed of four nodes.  

The measures we have proposed are ratios of the uniquely counted, closed motifs of order-four to the unique count of all connected structures of four nodes. This is conceptually convenient but these values may not be
suitable for use in low-dimensional ODE approximations, such as the pairwise model. Global clustering is not defined using purely unique counts (see appendix \ref{sec:unique}) 
and yields a different value when the unique counts are used. In appendix \ref{sec:unique}, we hypothesise the correct counts of motifs and paths for use
in clustering-type ratios. Whilst counting uniquely significantly reduces computational complexity, it has the slight disadvantage that it does not provide the multiplicative type of counting used in pairwise models. In the Appendix, we conjecture that this can be easily overcome by simply multiplying unique counts with the cardinality of the automorphism group corresponding to the motif.    

It has been demonstrated that care needs to be taken when trying to extend 
modelling to clustered networks. While models for simple clustered networks composed of exclusively non-overlapping triangles and edges have been developed, 
it is going to be more challenging to extend to networks with more complex structures and motifs. 
Motifs such as a square with a diagonal or a fully connected square may fulfil some function depending on the area of application (e.g. genetic regulatory networks, cortical networks), 
and thus measuring and quantifying this correctly is crucial for further model development. 
Many natural extensions for this work exist which include considerations around higher-order structure, algorithm efficiency in measuring these and development of synthetic network models 
that allow robust and transparent control of not only lower, but also higher-order structures.
%%%%%%%%%%%%%%%%%%%%%%%%%%%%%%%%%%%%%%%%%%%%%%%%%%%%%%%%%%%%%%%%%%%%%%%%%%%%%%%%%%%%%%%%%%%%%%%%%%

%%%%%%%%%%%%%%%
\section*{Acknowledgements}
%%%%%%%%%%%%%%%
Martin Ritchie acknowledges funding for his PhD studies from EPSRC (Engineering and Physical Sciences Research Council) and the University of Sussex. Thomas House is supported by the EPSRC, and would like to thank Charo I. Del Genio for discussions on the Motif Decomposition algorithm.

%%%%%%%%%%
\section{Appendix}
%%%%%%%%%

\subsection{Motif decomposition, analysis}\label{sec:mdanalysis}
%%%%%%%%%%%%%%
It is possible to write down the dynamics for this process in the limit of large networks by decomposing motifs into hyper nodes (considering each motif at a higher level as a node) and considering
the links between them. By identifying a hyper-node with $n$ edges as $Q_n$, it is possible to write equations for the normalised count of each hyper-node:

\begin{eqnarray}
 \frac{d}{dt}Q_5 &=& -6\lambda Q_5, \\
 \frac{d}{dt}Q_4 &=&  6\lambda Q_5 - 5\lambda Q_4, \\
 \frac{d}{dt}Q_3 &=&   \lambda Q_4 - 4\lambda Q_3, \\
 \frac{d}{dt}Q_2 &=&  4\lambda Q_3 - 3\lambda Q_2, \\
 \frac{d}{dt}Q_1 &=&  4\lambda Q_4 + 16\lambda Q_3 + 9\lambda Q_2.
\end{eqnarray}
These equations can be solved for initial condition $Q(0) = (0, 0, 0, 0, 1/4)$ and evolve
towards $Q(\infty) = (1, 0, 0, 0, 0)$. The rate $\lambda$ is just included for clarity and can be set to 1
with no loss of generality. The process stops at a `time' $t^*$ when the desired level of clustering has been  
achieved:
\begin{eqnarray}
 \frac{1}{6}\sum_iT_iQ_i(t^*) = \phi,
\end{eqnarray}
where $T_i$ denotes the number of triangles associated with each hyper-node type. The above equation may be
solved for $t^*$, and inserted into the the equations $(1)$-$(5)$ to obtain a prediction for motif structure. 
Such hyper-graph counting can be done for any $n$ but quickly becomes too tedious. It is also possible to use the quantities in Table 1 to
derive epidemic final sizes and other attributes.

\begin{minipage}{0.9\textwidth}
\begin{center}
\begin{tabular}[ht]{ r | r r r r r }
  $i$        & 1 & 2 & 3 &  4 &  5 \\
  $n_i$      & 1 & 3 & 4 &  4 &  4 \\
  $l_i$      & 0 & 6 & 8 & 10 & 12 \\
  $\sigma_i$ & 3 & 3 & 4 &  2 &  0 \\
  $T_i$      & 0 & 6 & 0 & 12 & 24 \\
\end{tabular}
\vspace{5mm}

Table 1: Table of hyper nodes indexed by $i$ with: the number of nodes involved $n_i$ , the
number of local links $l_i$ , the number of global stubs $\sigma_i$ , and the number of triangles $T_i$ .
\end{center}
\end{minipage}

\subsection{Motif counting algorithm}\label{sec:algorithm}
Below we introduce some notation in order to describe correctly and un-ambiguously the counting algorithm.

\begin{description}
  \item[Path] A path $P$ is an ordered tuple $(i,\dots,j)$.
  \item[$P_n$] The $n^{th}$ node of path $P$.
  \item[$H^n$] The head operator such that $H^n(P)$ returns the $n$ first nodes of the path $P$.
  \item[$T^n$] The tail operator such that $T^n(P)$ returns the $n$ last nodes of the path $P$.
  \item[$R$  ] The reverse operator such that $R((i,\dots,k,\dots,j))=(j,\dots,k,\dots,i)$.
  \item[$CP(P)$] The set of circular permutations of path $P$. 
  \item[$RCP(P)$] The set of all reverse circular permutations of path $P$ (\textbf{NB}: $RCP(P)\neq CP(P)$). 
  \item [$A$] The adjacency matrix, $A = A^T$ and with $Tr(A)=0$. 
  \item [$\{\cdot\}$] A set. 
  \item [$PLl$] The set of non-trivial paths of length $l$ ($l$ = number of edges). 
\end{description}

\begin{algorithm}[H]
/* {\em The following process is applied iteratively to determine non-trivial paths (open paths or closed paths) of length $l+1$ given non-trivial (non-loop) paths of length $l$. 
  The description below is not specific to a single length but assumes} $l\geq 2$. 
 \SetAlgoLined
 \KwData{The uniquely counted set of paths of length 2 is: $PL_2=\{(i,j,k)\}: A[i,k]\cdot A[k,j]>0$ and $j>i$.}
 initialization\;
  $LL_{l+1}=\emptyset$; /* {\em Closed paths of length} $l+1$ \\
  $PL_{l+1}=\emptyset$; /* {\em Open paths of length} $l+1$ \\
 \For{ All paths $P$ in $PL_l$}{
   \For{ All nodes $n$: $A[T^1(P),n]>0~ \& \notin T^1(P)$}{
	  $nP = (P,n)$; /* {\em new path} \\
  \If{$n=H^1(L)~\&~ nP \notin LL_{l+1} ~\&~ R(nP)\notin LL_{l+1}$}{
    $LL_{l+1} \leftarrow nP$\;} 
  \If{$n\neq H^1(L)~\&~ nP \notin LL_{l+1} ~\&~ R(nP)\notin LL_{l+1}$}{
    $PL_{l+1} \leftarrow nP$\;}
     /* {\em These exclude symmetric paths but not circular permutations.} 
      }
 \If{ $P^* \in \{ CP(LL_{l+1}),CP(PL_{l+1}) \}$}{
      \{$LL_{l+1}\} =\{LL_{l+1} \} \backslash P^* $;\\
      \{$PL_{l+1}\} =\{LL_{l+1} \} \backslash P^* $;}
/* {\em Removes circular permutations.}
}
 \For{ All paths $P \in CLL_{l+1}$}{
       \If{$P^* \in CLL_{l+2}~\&~P^* \notin CRP(CLL_{l+1}) $}{
            $LL_{l+1} \leftarrow P$;}
 }
 \For{ All paths $P \in CPL_{l+1}$}{
       \If{$P^* \in CPL_{l+2}~\&~P^* \notin CRP(CPL_{l+1}) $}{
            $PL_{l+1} \leftarrow P$;}
 }
/* {\em Removes reverse circular permutations.}
 \caption{Pseudo code for the motif counting algorithm.}
\end{algorithm}

\subsection{Motif counting: unique vs multiplicative}\label{sec:unique}

In this paper all order-four clustering-type ratios use unique counts. Ratios based on unique counts will give different values to ratios based on multiplicative counts. As an example of multiplicative counting, classic clustering is defined as:
\begin{eqnarray}
 \phi = \frac{6 \times [\triangle]}{[\triangle + \vee]},
\end{eqnarray}
where $[\triangle]$ denotes the number of triangles, and $[\triangle + \vee]$ the number of length three paths closed and unclosed (doubly counted) 
in the network. If unique counts are used then we have $\phi_{unique} = \phi/3$. 

We have computed the unique order-four counts in order to improve the computational performance of our algorithm.
However, if we wish to normalise or scale-up the unique counts to correspond to the multiplicative equivalent, correct multiplying factors need to be determined. This appears to be the number of automorphisms associated with each motif type or path length: a triangle has six and a path of length three has two automorphisms. 

Let $A = (a_{i,j}),~i,j \in \{1,\dots N \},$ be the adjacency matrix of an undirected network with no self loops i.e. $A=A^T$ and $A_{i,i}=0$ for any $i = 1,2,\dots,N$. 
It is possible to obtain the multiplicative counts from the adjacency matrix $A$. Summing over all nodes:
\begin{eqnarray}
 [-]= \sum_{i,j=1,~i \neq j}^N a_{i,j}.
\end{eqnarray}
This counts twice the number of real or uniquely counted edges in the network. It is possible to count more complex paths as well:

 \begin{eqnarray}
  [\vee + \triangle] = \sum_{i,j,k=1,~i\neq j \neq k}^N a_{i,j}a_{j,k},
 \end{eqnarray}
yielding all connected structures of 3 nodes (closed and unclosed), similarly
 \begin{eqnarray}
[\triangle] = \sum_{i,j,k=1,~i\neq j \neq k}^N a_{i,j}a_{i,k}a_{j,k},
 \end{eqnarray}
yielding six times the number of unique triangles. It is also possible to count six different closed paths of length three contained within a triangle: for each node
we count clockwise and counter-clockwise about the triangle. This by-directional counting is important so that the method is consistent when considering directed networks. 
Following the same counting methodology it is possible to count order-four structures:

\begin{eqnarray}
 [\sqcup + \boxempty + \boxslash + \boxtimes] = \sum_{i,j,k,l=1~i\neq j \neq k \neq l}^N a_{ij}a_{jk}a_{kl}.
\end{eqnarray}
In this form we see that it is possible to compute the individual counts using the following identities:
\begin{eqnarray}
 [\boxtimes] &=& \sum_{i,j,k,l=1,~i\neq j \neq k \neq l}^N a_{ij}a_{ik}a_{il}a_{jk}a_{jl}a_{kl}, 
\end{eqnarray}
\begin{eqnarray}
 [\boxslash] &=& \sum_{i,j,k,l=1,~i\neq j \neq k \neq l}^N a_{ij}a_{ik}(1-a_{il})a_{jk}a_{jl}a_{kl}, 
\end{eqnarray}
\begin{eqnarray}
 [\boxempty] &=& \sum_{i,j,k,l=1,~i\neq j \neq k \neq l}^N a_{ij}a_{ik}(1-a_{il})(1-a_{jk})a_{jl}a_{kl},
\end{eqnarray}
Counting this way a single $[\boxtimes]$ is counted 24 times, $ [\boxslash]$ is counted 4 times and $[\boxempty]$ is counted 8 times, equal to the number
of automorphisms associated with each motif type. Currently, based on our intuition and numerical tests, we conjecture that this is the correct way to scale-up from unique to multiplicative motif counts. 
This method of counting is thorough but not practical for networks of reasonable size since it has complexity $\mathcal{O}(N^n)$ for order-$n$ structures.
\bibliography{sample}{}
 \bibliographystyle{ieeetr}
\nocite{*}

\newpage

\newpage

\newpage

\newpage

%%%%%%%%%%%%%%%%%%%%%%%%%%%%%%%%%%%%%%%%%%%%%%%%%%%%%%%%%%%%%%%%%%%%%%%%%%%%%%%%%%%%%%%%%%%%%%%%%%%%%%%%%%%%%%%%%%%%%%%%%%%%%%%%%%%%%%

\newpage

\newpage

%%%%%%%%%%%%%%%%%%%%%%%%%%%%%%%%%%%%%%%%%%%%%%%%%%%%%%%%%%%%%%%%%%%%%%%%%%%%%%%%%%%%%%%%%%%%%%%%%%%%%%%%%%%%%%%%%%%%%%%%%%%%%%%%%%%%%%%
% Both sets of boxplots
\newpage

\newpage
%%%%%%%%%%%%%%%%%%%%%%%%%%%%%%%%%%%%%%%%%%%%%%%%%%%%%%%%%%%%%%%%%%%%%%%%%%%%%%%%%%%%%%%%%%%%%%%%%%%%%%%%%%%%%%%%%%%%%%%%%%%%%%%%%%%%%%%
% Edge percolation plots
\newpage

\newpage

\begin{figure}
\begin{center}
\includegraphics[scale=0.5]{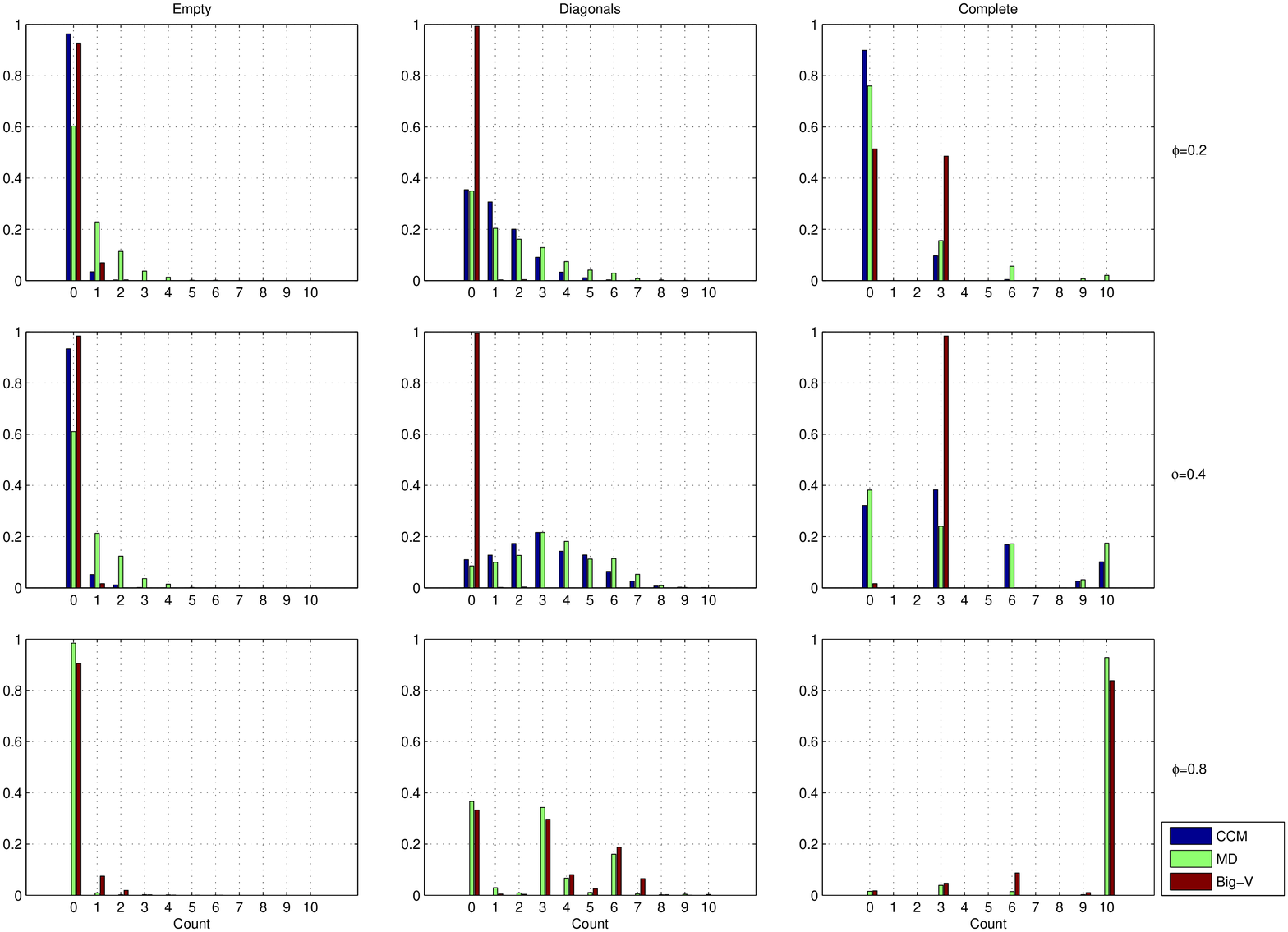}
\captionof{figure}{Order-four motif distribution. The per-node distribution of the number of unique counts of order-four motifs, for all previously used networks. See Appendix~\ref{sec:algorithm} which details how motifs are counted.}
\label{fig:motifdist}
\end{center} 	
\end{figure}
%%%%%%%%%%%%%%%%%%%%%%%%%%%%%%%%%%%%%%%%%%%%%%%%%%%%%%%%%%%%%%%%%%%%%%%%%%%%%%%%%%%%%%%%%%%%%%%%%%%%%%%%%%%%%%%%%%%%%%%%%%%%%%%%%%%%%%
\newpage

\newpage
%%%%%%%%%%%%%%%%%%%%%%%%%%%%%%%%%%%%%%%%%%%%%%%%%%%%%%%%%%%%%%%%%%%%%%%%%%%%%%%%%%%%%%%%%%%%%%%%%%%%%%%%%%%%%%%%%%%%%%%%%%%%%%%%%%%%%%
\newpage

\newpage
%%%%%%%%%%%%%%%%%%%%%%%%%%%%%%%%%%%%%%%%%%%%%%%%%%%%%%%%%%%%%%%%%%%%%%%%%%%%%%%%%%%%%%%%%%%%%%%%%%%%%%%%%%%%%%%%%%%%%%%%%%%%%%%%%%%%%%

%%%%%%%%%%%%%%%%%%%%%%%%%%%%%%%%%%%%%%%%%%%%%%%%%%%%%%%%%%%%%%%%%%%%%%%%%%%%%%%%%%%%%%%%%%%%%%%%%%%%%%%%%%%%%%%%%%%%%%%%%%%%%%%%%%%%%%

%%%%%%%%%%%%%%%%%%%%%%%%%%%%%%%%%%%%%%%%%%%%%%%%%%%%%%%%%%%%%%%%%%%%%%%%%%%%%%%%%%%%%%%%%%%%%%%%%%%%%%%%%%%%%%%%%%%%%%%%%%%%%%%%%%%%%%
\newpage

% \begin{figure}
%  \begin{center} 
% \begin{tabular}{cc}
%  \includegraphics[scale=0.55]{gccclust} & 
%  \includegraphics[scale=0.55]{gccboxc04} \\
%  \includegraphics[scale=0.55]{gccboxc05} &
%  \includegraphics[scale=0.55]{gccboxc06} \\
% \end{tabular}
%  \end{center}
% \caption{The effect of clustering on GCC size. As the rewiring algorithms worked the networks the size of the GCC was recorded for varying values of $\phi$. The Big-V data ranges from $0.05\leq \phi_v \leq 0.85$, 
% it becomes computationally expensive for the algorithm to push clustering much higher than 0.85. Similarly the MD algorithm data ranges from $0.1 \leq \phi_{md} \leq 0.9$, this algorithm takes time to fully 
% reconstruct structure into a more random network-like arrangement. The boxplots were chosen where the disparity in GCC for the same levels of clustering was greatest.}
% \label{fig:gccclust}
% \end{figure}
% 
% \newpage
%%%%%%%%%%%%%%%%%%%%%%%%%%%%%%%%%%%%%%%%%%%%%%%%%%%%%%%%%%%%%%%%%%%%%%%%%%%%%%%%%%%%%%%%%%%%%%%%%%%%%%%%%%%%%%%%%%%%%%%%%%%%%%%%%%%%%%

 \end{document}